\documentclass[journal=jctc,manuscript=article]{achemso}
\usepackage[utf8]{inputenc}
\usepackage[T1]{fontenc}
\usepackage{xcolor}
\usepackage{amsmath}
\usepackage{amssymb}
\usepackage{graphicx}
\usepackage{multirow}
\usepackage[version=4]{mhchem}
\usepackage[section]{placeins}
\usepackage{float}
\usepackage{booktabs}
\usepackage{silence}
\usepackage{array}
\usepackage{braket}
\newcolumntype{K}{>{\centering\arraybackslash}b{0.85cm}}
\newcolumntype{L}{>{\centering\arraybackslash}b{1.28cm}}
\usepackage[english]{babel}
\usepackage{subfigure}
\usepackage{microtype}
\SectionNumbersOn



\title{Benchmark of \textit{GW} Methods for Core-Level Binding Energies}
\author{Jiachen Li}
\affiliation{Department of Chemistry, Duke University, Durham, NC 27708, USA}
\author{Ye Jin}
\affiliation{Department of Chemistry, Duke University, Durham, NC 27708, USA}
\author{Patrick Rinke}
\affiliation{Department of Applied Physics, Aalto University, Otakaari 1, FI-02150 Espoo, Finland}
\author{Weitao Yang}
\affiliation{Department of Chemistry, Duke University, Durham, NC 27708, USA}
\author{Dorothea Golze}
\affiliation{Faculty of Chemistry and Food Chemistry, Technische Universit\"at Dresden, 01062 Dresden, Germany}
\alsoaffiliation{Department of Applied Physics, Aalto University, Otakaari 1, FI-02150 Espoo, Finland}
\email{dorothea.golze@tu-dresden.de}

\begin{document}

\begin{tocentry}
\includegraphics[width=1\textwidth]{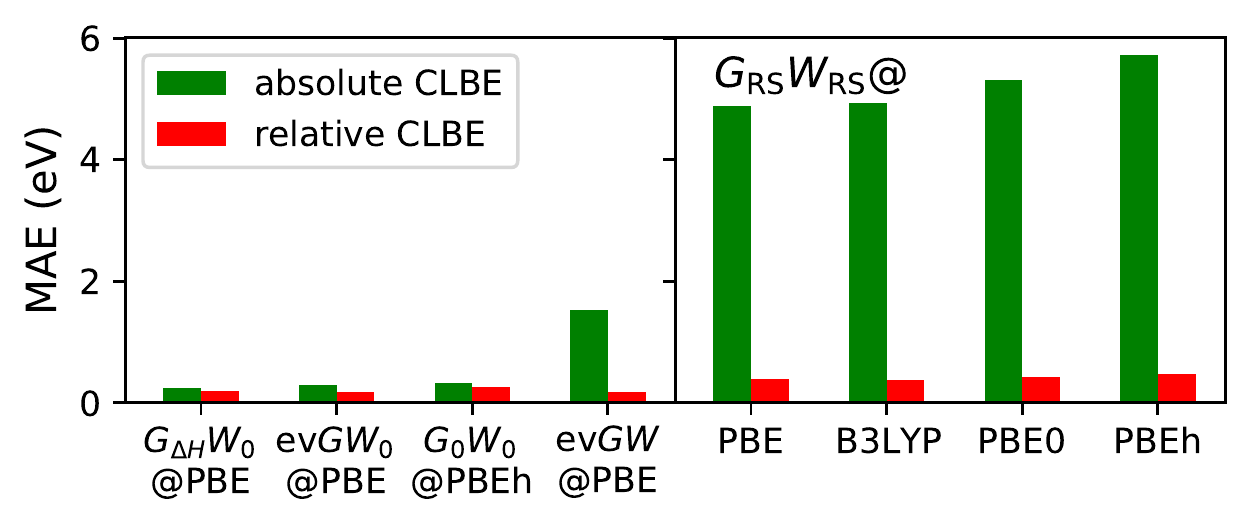}
\end{tocentry}

\begin{abstract}
The $GW$ approximation has recently gained increasing attention as a viable method for the computation of deep core-level binding energies as measured by X-ray photoelectron spectroscopy (XPS). We present a comprehensive benchmark study of different $GW$ methodologies (starting-point optimized, partial and full eigenvalue-self-consistent, Hedin shift and renormalized singles) for molecular inner-shell excitations. We demonstrate that all methods yield a unique solution and apply them to the CORE65 benchmark set and ethyl trifluoroacetate. Three $GW$ schemes clearly outperform the other methods for absolute core-level energies with a mean absolute error of 0.3~eV with respect to experiment. These are  partial eigenvalue self-consistency, in which the eigenvalues are only updated in the Green's function, single-shot $GW$ calculations based on an optimized hybrid functional starting point and a Hedin shift in the Green's function. While all methods reproduce the experimental relative binding energies well, the eigenvalue self-consistent schemes and the Hedin shift yield with mean errors $<0.2$~eV the best results.
\end{abstract}
\maketitle

\section{INTRODUCTION}\label{sec:introduction}
X-ray photoelectron spectroscopy (XPS) is a standard characterization tool for  materials,\cite{bagusInterpretationXPSSpectra2013} liquids\cite{hofftElectronicStructureSurface2006,j.villar-garciaChargingIonicLiquid2011} and molecules.\cite{siegbahnESCAAppliedFree1969} In XPS, the core-level binding energies (CLBEs) are measured,\cite{barrModernESCAPrinciples2020} which are element-specific, but also sensitive to the chemical environment. However, establishing the link between between the measured spectrum and the atomic structure is challenging, in particular for complex materials with heavily convoluted XPS signals.\cite{aarvaUnderstandingXraySpectroscopy2019,aarvaUnderstandingXraySpectroscopy2019a} Guidance from theory is thus often necessary to interpret XPS spectra.

Theoretical XPS methods can be distinguished into Delta and response approaches. In Delta-based methods, the CLBE is computed as the total energy difference between the neutral and core-excited system. These calculations can be performed at different levels of theory, for example, with high-level wave-function methods, such as Delta coupled cluster ($\Delta$CC) \cite{zhengPerformanceDeltaCoupledClusterMethods2019,senInclusionOrbitalRelaxation2018} or with Kohn-Sham density functional theory\cite{hohenbergInhomogeneousElectronGas1964,kohnSelfConsistentEquationsIncluding1965,parrDensityFunctionalTheoryAtoms1989} (KS-DFT). The most popular DFT-based approach is the Delta self-consistent field ($\Delta$SCF) method,\cite{bagusSelfConsistentFieldWaveFunctions1965} which has been thoroughly benchmarked. \cite{pueyobellafontPredictionCoreLevel2015,pueyobellafontPerformanceMinnesotaFunctionals2016,vinesPredictionCoreLevel2018,kahkAccurateAbsoluteCoreelectron2019,kleinNutsBoltsCorehole2021,kahkCoreElectronBinding2021} While high accuracy can be achieved with these approaches, the explicit optimization of a core-ionized wave function leads to conceptual problems, e.g., regarding periodicity, constraining spin-orbit coupled states or, in the case of DFT, deteriorating accuracy for larger structures, which was already discussed and demonstrated elsewhere.\cite{pinheiroLengthDependenceIonization2015,golzeCoreLevelBindingEnergies2018,michelitschEfficientSimulationNearedge2019,golzeAccurateComputationalPrediction2021,hallSelfinteractionErrorInduces2021}

An explicit orbital optimization of core-ionized systems and the related conceptual issues are avoided in response theories, where electron propagators are applied to transform the ground into an excited state. Recently, wave-function-based methods, such as linear response and equation-of-motion CC methods\cite{liuBenchmarkCalculationsKEdge2019,vidalEquationofMotionCoupledClusterTheory2020,ambroiseProbingBasisSet2021,vilaEquationofMotionCoupledClusterCumulant2021} and the algebraic diagrammatic construction method\cite{ambroiseProbingBasisSet2021} were reassessed for absolute CLBEs, yielding partly promising results. Another promising approach in the realm of response methods is the $GW$ approximation\cite{martinInteractingElectrons2016,golzeGWCompendiumPractical2019,reiningGWApproximationContent2018} to many-body perturbation theory, which is derived from Hedin's equation\cite{hedinNewMethodCalculating1965} by omitting the vertex correction. The $GW$ approximation is considered the ``gold standard" for the computation of band structures of materials,\cite{golzeGWCompendiumPractical2019,rasmussenFullyAutomatedGW2021} but it has also been successfully applied to valence excitations of molecules.\cite{golzeGWCompendiumPractical2019,vansettenGW100BenchmarkingG0W02015,stukeAtomicStructuresOrbital2020,brunevalGWMiracleManyBody2021} 

Due to its primary application to solids, $GW$ was traditionally implemented in plane wave codes that typically use pseudo potentials for the deeper states. With the increasing availability of the $GW$ method in localized basis set codes,\cite{blaseFirstprinciplesMathitGWCalculations2011, renResolutionofidentityApproachHartree2012, vansettenGWMethodQuantumChemistry2013,brunevalMolgwManybodyPerturbation2016, wilhelmGWGaussianPlane2016,forsterQuadraticPairAtomic2020,zhuAllElectronGaussianBasedG0W02021,mejia-rodriguezScalableMolecularGW2021} core states moved into focus. Core-level binding energy calculations have emerged as a recent trend in $GW$. \cite{aokiAccurateQuasiparticleCalculation2018,vansettenAssessingGWApproaches2018,golzeCoreLevelBindingEnergies2018,vooraEffectiveOneparticleEnergies2019,golzeAccurateAbsoluteRelative2020,kellerRelativisticCorrectionScheme2020,ducheminRobustAnalyticContinuationApproach2020,zhuAllElectronGaussianBasedG0W02021,mejia-rodriguezScalableMolecularGW2021,mejia-rodriguezBasisSetSelection2022,golzeAccurateComputationalPrediction2021} By extension to the Bethe-Salpeter equation (BSE@$GW$) also $K$-edge transition energies measured in X-ray absorption spectroscopy can be calculated.\cite{yaoAllElectronBSEGW2022} These studies focused primarily on molecules. However, $GW$ is one of the most promising methods for core-level predictions of materials because the scaling with respect to system size is generally smaller than for wave-function based response methods and the method is well-established for periodic structures. In addition, $GW$ implementations for localized basis sets are advancing rapidly. Recently, periodic implementations\cite{wilhelmPeriodicGWCalculations2017,renAllelectronPeriodic0W2021,zhuAllElectronGaussianBasedG0W02021} and low-scaling $GW$ algorithms with $O(N^3)$ complexity emerged in localized basis sets formulations.\cite{wilhelmGWCalculationsThousands2018,forsterQuadraticPairAtomic2020,wilhelmLowScalingGWBenchmark2021,ducheminCubicScalingAllElectronGW2021,forsterLowOrderScalingQuasiparticle2021}  

The application of $GW$ to core states is more difficult than for valence states. We showed that more accurate and computationally more expensive techniques for the frequency integration of the self-energy are required compared to valence excitations.\cite{golzeCoreLevelBindingEnergies2018} Furthermore, we found that the standard single-shot $G_0W_0$ approach  performed on top of  DFT calculations with generalized gradient approximations (GGAs) or standard hybrid functionals fails to yield a distinct solution, which is caused by a loss of spectral weight in the quasiparticle (QP) peak.\cite{golzeAccurateAbsoluteRelative2020} We demonstrated that eigenvalue self-consistency in $G$ or using a hybrid functional with 45\% exact exchange as starting point for the $G_0W_0$ calculation restores the QP main excitation. Including also relativistic corrections,\cite{kellerRelativisticCorrectionScheme2020} an agreement of 0.3~eV and 0.2~eV with respect to experiment was reported for absolute and relative CLBEs, respectively.\cite{golzeAccurateAbsoluteRelative2020}

While $G_0W_0$ is the computationally least expensive $GW$ flavor, it strongly depends on the density functional approximation (DFA). Tuning the exchange in the hybrid functional to, e.g., 45\% is conceptually unappealing and introduces undesired small species dependencies, as discussed more in detail in this work. Self-consistency reduces or removes the dependence on the underlying DFT functional, but significantly increases the computational cost. The computationally least expensive self-consistent schemes are the so-called eigenvalue self-consistent approaches, where the eigenvalues are iterated in the Green's function $G$ (ev$GW_0$) or alternatively in $G$ and the screened Coulomb interaction $W$ (ev$GW$). 
\cite{golzeGWCompendiumPractical2019} Higher-level self-consistency schemes, such as fully-self-consistent $GW$\cite{schoneSelfConsistentCalculationsQuasiparticle1998,carusoUnifiedDescriptionGround2012,carusoSelfconsistentGWAllelectron2013} (sc$GW$) and quasiparticle self-consistent $GW$\cite{vanschilfgaardeQuasiparticleSelfConsistentGW2006} (qs$GW$) remove, unlike ev$GW_0$ or ev$GW$, the starting point dependence completely. However, these higher-level self-consistency schemes are much more expensive and not necessarily better because of the inherent underscreening due to the missing vertex correction.\cite{carusoBenchmarkGWApproaches2016,grumetQuasiparticleApproximationFully2018}


Recently, we proposed the renormalized singles (RS) Green's function approach, denoted as $G_{\text{RS}}W_0$, to reduce the starting point dependence in $GW$.\cite{jinRenormalizedSinglesGreen2019}
The RS concept was developed in the context of the random phase approximation (RPA) for accurate correlation energies,\cite{renRandomPhaseApproximationElectron2011,renRenormalizedSecondorderPerturbation2013}
and termed renormalized single-excitation (RSE) correction.
Following standard perturbation theory, single-excitations  contribute to the second-order correlation energy. It was shown that their inclusion significantly improves binding energies.\cite{renRandomPhaseApproximationElectron2011,renRenormalizedSecondorderPerturbation2013}
The RS Green's function approach extends the RSE idea from correlation energies to $GW$ QP energies.
In the $G_{\mathrm{RS}}W_0$ scheme, the RS Green's function is used as a new starting point and the screened Coulomb interaction is calculated with the KS Green's function. For valence excitations, we found that this renormalization process significantly reduces the starting point dependence and provides improved accuracy over $G_0W_0$.\cite{jinRenormalizedSinglesGreen2019} The mean absolute errors obtained from the $G_{\text{RS}}W_{0}$ approach with different DFAs are smaller than $0.2$ \,{eV} for predicting ionization potentials of molecules in the $GW$100 set.\cite{jinRenormalizedSinglesGreen2019}  Unlike the self-consistent schemes, the RS Green's function method hardly increases the computational cost compared to $G_0W_0$.\par  

Recently, we employed the concept of RS in a multireference DFT approach for strongly correlated systems.\cite{liMultireferenceDensityFunctional2022} We also used the RS Green's function in the T-matrix approximation ($G_{\mathrm{RS}}T_{\mathrm{RS}}$) in a similar manner as in $G_{\text{RS}}W_{\text{RS}}$.\cite{liRenormalizedSinglesGreen2021} The T-matrix method scales formally as $\mathcal{O}(N^6)$ with respect to system size $N$,\cite{liRenormalizedSinglesGreen2021,zhangAccurateQuasiparticleSpectra2017} with 
reduced scaling possible using effective truncation of the active space.\cite{zhangAccurateEfficientCalculation2016} 
In addition to the high computational cost, the performance of $G_{\mathrm{RS}}T_{\mathrm{RS}}$ for core-levels is not particularly impressive. The error with respect to experiment is 1.5~eV for absolute and 0.3~eV for relative CLBEs.\cite{liRenormalizedSinglesGreen2021}
In the present work, we focus thus on RS $GW$ approaches for core-levels.

In this work, we benchmark $GW$ approaches, which we consider computationally affordable and suitable for large-scale applications. This includes $G_0W_0$ with tuned starting points, the  eigenvalue self-consistent schemes ev$GW_0$ and ev$GW$ and two new methods that we introduce in this work. One is based on the so-called Hedin shift\cite{pollehnAssessmentLessGreater1998} and can be understood as approximation of the ev$GW_0$ method. We refer to this scheme as $G_{\Delta\mathrm{H}}W_0$. The other is a different flavor of the RS Green's function approach, where the the screened Coulomb interaction is also computed with the RS Green's function ($G_{\mathrm{
RS}}W_{\mathrm{RS}}$). 

The remainder of this article is organized as follows: We introduce the different $GW$ approaches in Section~\ref{sec:theory} and give the computational details for our calculations in Section~\ref{sec:computation}. The solution behavior of the different methods is discussed in Section~\ref{subsec:graphical_solutions} by comparing self-energy matrix elements and spectral functions. In Section~\ref{subsec:core65} results are presented for the CORE65 benchmark set and in Section~\ref{subsec:etfa} for the ethyl trifluoroacetate molecule and we finally draw conclusions in Section~\ref{sec:conclusion}.

\section{\label{sec:theory}THEORY}

\subsection{Single-shot $\boldsymbol{G_0W_0}$ approach}
\label{subsec:g0w0}
The most popular $GW$ approach is the single-shot $G_0W_0$ scheme, where the $GW$ QP energies are obtained as corrections to the KS eigenvalues $\{\epsilon^0_n\}$: 
\begin{equation}
\epsilon_{n}^{\text{QP}} = \epsilon_{n}^0 + \mathrm{Re}\braket{\psi_{n}^0|\Sigma(\epsilon_{n}^{\text{QP}})- v^{\mathrm{xc}}|\psi_{n}^0}. 
\label{eq:qpequation}
\end{equation}
$\{\psi_n^0\}$ are the KS molecular orbitals (MOs) and $v^{xc}$ is the KS exchange-correlation potential. 
We use $i$, $j$ for occupied orbitals and $a$, $b$ for virtual orbitals, and $m$, $n$ for general orbitals. We omitted the spin index in all equations for simplicity and use the notation  $\Sigma_{n}=\braket{\psi_{n}^0|\Sigma|\psi_{n}^0}$ and $v^{\mathrm{xc}}_n=\braket{\psi_{n}^0|v^{\mathrm{xc}}|\psi_{n}^0}$ in the following. We can directly obtain the CLBE of state $n$ from the QP energies because they are related by $\text{CLBE}_n= -\epsilon_{n}^{\text{QP}}$. The self-energy $\Sigma$ is given by
\begin{equation}
 \Sigma(\mathbf{r},\mathbf{r}',\omega) = \frac{i}{2\pi}\int \mathrm{d}\omega' G_0(\mathbf{r},\mathbf{r}',\omega+\omega')W_0(\mathbf{r},\mathbf{r}',\omega')e^{i\omega'\eta}
 \label{eq:Sigma}
\end{equation}
where the non-interacting KS Green's function is denoted $G_0$ and the screened Coulomb interaction $W_0$. $\eta$ is a positive infinitesimal. The KS Green's function reads
\begin{equation}
 G_0(\mathbf{r},\mathbf{r}',\omega) = \sum_m\frac{\psi_{m}^0(\mathbf{r})\psi_{m}^0(\mathbf{r'})}{\omega - \epsilon_{m}^0 - i\eta \mathrm{sgn}(\epsilon_{\mathrm{F}} - \epsilon_{m}^0)}
 \label{eq:G0}
\end{equation}
where $\epsilon_{\mathrm{F}}$ is the Fermi energy.
%
The screened Coulomb interaction is calculated at the level of the RPA as
\begin{equation}
 W_0(\mathbf{r},\mathbf{r}',\omega) = \int \mathrm{d}\mathbf{r}'' \varepsilon^{-1}(\mathbf{r},\mathbf{r}'',\omega)v(\mathbf{r}'',\mathbf{r}')
\end{equation}
where $\varepsilon(\mathbf{r},\mathbf{r}',\omega)$ is the dielectric function and $v(\mathbf{r},\mathbf{r}') = 1/|\mathbf{r}-\mathbf{r}'|$ the bare Coulomb interaction. 

The calculation of the self-energy matrix elements $\Sigma_{n}$ is split into a correlation part $\Sigma^c$ and an exchange part $\Sigma^x$, i.e., $\Sigma_{n} = \Sigma_{n}^c + \Sigma_{n}^x$. The HF-like exchange part $\Sigma^x_n$ is given by
\begin{equation}
    \Sigma^x_n = -\sum_i^{\text{occ}} \Braket{\psi_n^0\psi_i^0|\psi_i^0\psi_n^0}.
 \label{eq:sigmax}
\end{equation}
The correlation part $\Sigma^c$ is computed from $W_0^c =W_0 -v$ and is the part that we plot in Section~\ref{subsec:graphical_solutions} to investigate the $GW$ solution behavior. The correlation  part in its fully analytic form is given by
\begin{equation}
 \Sigma_n^c(\omega) = \sum_m \sum_{s} \frac{| \Braket{\psi_n^0\psi_m^0| \rho^0_s} |^2 }{\omega -\epsilon_m^0 + 
(\Omega_s^0-i\eta)\mathrm{sgn}(\epsilon_{\mathrm{F}} -\epsilon_m^0)},
    \label{eq:g0w0sigma_pole}
\end{equation}
where $\Omega_s^0$ are charge neutral excitations at the RPA level and $\rho^0_s$ are the corresponding transition densities.
The fully analytic form of $\Sigma_n^c$ directly shows the pole structure of the self-energy and is illustrative to understand the solution behavior of $GW$. However, the evaluation of $\Omega_s^0$ scales with $O(N^6)$. In practice, the correlation self-energy is usually evaluated with a reduced scaling by using techniques such as analytical continuation\cite{ducheminRobustAnalyticContinuationApproach2020,zhuAllElectronGaussianBasedG0W02021} or the contour deformation.\cite{govoniLargeScaleGW2015,golzeCoreLevelBindingEnergies2018,zhuAllElectronGaussianBasedG0W02021}\par
The QP energies can be obtained by solving Equation~\eqref{eq:qpequation}, which is typically the computationally least expensive approach. In this work, we additionally employ two alternative approaches to obtain further insight into the physics and suitability of the different $GW$ approaches. The first is the graphical solution of Equation~\eqref{eq:qpequation}, where we plot the self-energy matrix elements $\Sigma^c_n$ and determine the QP solution by finding the intersections with the straight line $\omega-\varepsilon_{n}^0+v_{n}^{\mathrm{xc}}-\Sigma_{n}^x$. The presence of several intersections would indicate that more than one solution exists. The second, computationally even more expensive alternative, is the computation of the spectral function,\cite{golzeAccurateAbsoluteRelative2020} which is given by
\begin{equation}
A(\omega)=\frac{1}{\pi}\sum_m\frac{\left|\text{Im}\Sigma_m(\omega)\right|}{\left[\omega-\epsilon_m-\left(\text{Re}\Sigma_m(\omega)-v_m^{\mathrm{xc}}\right)\right]^2+\left[\text{Im}\Sigma_m(\omega)\right]^2}
 \label{eq:Aomegatrace}
\end{equation}
%
In Equation~\eqref{eq:Aomegatrace}, we include also the imaginary part of the complex self-energy, which gives us direct access to the spectral weights and satellite spectrum. At the $G_0W_0$  level, we only use the diagonal elements of $\Sigma$ in Equation~\eqref{eq:Aomegatrace}.

\subsection{Eigenvalue-selfconsistent \textit{GW} schemes}

Including self-consistency in Hedin's $GW$ equations is a widely used strategy to go beyond $G_0W_0$. sc$GW$\cite{golzeGWCompendiumPractical2019,carusoSelfconsistentGWAllelectron2013} is conceptually the purest approach, but also the most expensive one. To reduce the computational demands, different lower-level self-consistent schemes were developed. The simplest approach is an eigenvalue self-consistent scheme, which comes in two different flavors. The first one is to iterate the eigenvalues only in $G$ and keep $W$ fixed at the $W_0$ level. This scheme is referred to as the ev$GW_0$ approach. In ev$GW_0$, we start by updating the KS eigenvalues in the Green's function with the $G_0W_0$ QP energies, re-evaluate the QP equation (see Equation~\eqref{eq:qpequation}) and iterate until self-consistency in $G$ is reached. The Green's function in the eigenvalue self-consistent scheme reads
\begin{equation}
 G_{\textnormal{ev}}(\mathbf{r},\mathbf{r}',\omega) =  \sum_m\frac{\psi_{m}^0(\mathbf{r})\psi_{m}^{0}(\mathbf{r}')}{\omega-\epsilon_{m}^{\text{QP}} - i\eta  \text{sgn}(\epsilon_{\rm F}-\epsilon_{m}^{\text{QP}})} \label{eq:evgreensfkt}
\end{equation}
with $\epsilon_m^{\text{QP}} = \epsilon_m^0 + \Delta \epsilon_m$, where $\Delta\epsilon_m$ is the $GW$ correction, see Equation~\eqref{eq:qpequation}. The second flavor is ev$GW$, where the KS eigenvalues are not only updated in $G$, but also in the screened Coulomb interaction $W$.
The eigenvalue self-consistent calculations are computationally significantly more expensive than a $G_0W_0$ calculation, in particular in combination with the accurate self-energy integration techniques that are required for core-levels.\cite{golzeCoreLevelBindingEnergies2018} The computational demands are large because $G$, the screened Coulomb interaction $W$ (in the case of ev$GW$) and the self-energy have to be built repeatedly. In addition, 
 Equation~\eqref{eq:qpequation} must be solved not only for the states of interest, but for all states.

\subsection{\textit{GW} with Hedin shift}
The cost of an ev$GW_0$ scheme can be drastically reduced by using a global shift $\Delta \mathrm{H}$ instead of an individual shift $\Delta \epsilon_m$ for each state $m$. This scheme was first introduced by Hedin~\cite{hedinNewMethodCalculating1965} and is referred to as $G_{\Delta\mathrm{H}}W_0$ in the following. The $G_{\Delta\mathrm{H}}W_0$ approach was discussed several times in the literature~\cite{leeTransitionAdiabaticSudden1999,pollehnAssessmentLessGreater1998,rinkeCombiningGWcalculationsExactexchangeDensityfunctional2005,martinInteractingElectrons2016} and the effect of ev$GW_0$ and the $G_{\Delta\mathrm{H}}W_0$ on the self-energy structure has been discussed for valence states in Ref.~\citenum{golzeGWCompendiumPractical2019}. In the Hedin-shift scheme, the Green's function transforms into
\begin{equation}
 G_{\Delta \mathrm{H}}(\mathbf{r},\mathbf{r}',\omega) =  \sum_m\frac{\psi_{m}^0(\mathbf{r})\psi_{m}^{0}(\mathbf{r}')}{\omega-(\epsilon_{m}^0+\Delta \text{H})- i\eta \text{sgn}(\epsilon_{\rm F}-\epsilon_{m}^0)},
 \label{eq:deltagreensfkt}
\end{equation}
where $G_0(\omega-\Delta\mathrm{H})=G_{\Delta \mathrm{H}}(\omega)$. The QP equation with the Hedin-shift scheme then becomes
\begin{equation}
\epsilon_{n}^{\text{QP}} = \epsilon_{n}^0 + \mathrm{Re}\braket{\psi_{n}^0|\Sigma(\epsilon_{n}^{\text{QP}}-\Delta {\rm H} )- v^{\mathrm{xc}}|\psi_{n}^0}. 
\label{eq:qpequationdeltaH}
\end{equation}

Traditionally, $\Delta$H is determined with respect to the Fermi level of $G_0$ for metals or the valence band maximum for gapped solid-state systems. For the molecular case, $\Delta$H is evaluated with respect to the highest occupied molecular orbital (HOMO) by introducing the self-consistency condition $\epsilon_{\rm{HOMO}}^{\rm QP} =  \epsilon^0_{\rm{HOMO}} + \Delta \rm{H}$, which is inserted in Equation~\eqref{eq:qpequationdeltaH} and yields
\begin{equation}
  \Delta \mathrm{H}_{\rm HOMO}  = \mathrm{Re} \Sigma_{\rm HOMO}(\epsilon_{\rm HOMO}^0)- v_{\rm HOMO}^{\mathrm{xc}}.
  \label{eq:Hshift}
\end{equation}
As demonstrated in Ref.~\citenum{golzeGWCompendiumPractical2019}, ev$GW_0$ and the $G_{\Delta \rm H}W_0$ lead to a shift of the pole structure of the self-energy matrix elements $\Sigma_n$ to more negative frequencies. For $\Sigma_{\text{HOMO}}(\omega)$, the shift is similar for ev$GW_0$ and $G_{\Delta \rm H}W_0$ yielding practically the same $\epsilon_{\rm HOMO}^{\text{QP}}$. 

For core states, we found that the shift $\Delta \rm{H}$ computed as in Equation~\eqref{eq:Hshift} is much smaller than the one from ev$GW_0$. We propose here a new approach, where we calculate the shift $\Delta \rm{H}$ for the core state of interest as
\begin{equation}
  \Delta \rm{H}_{\rm core} = \mathrm{Re}\Sigma_{\mathrm{core}}(\epsilon_{\rm core}^0)- v_{\mathrm{core}}^{\mathrm{xc}}.
  \label{eq:Hshiftcore}
\end{equation}
For example, to obtain $\epsilon^{\rm QP}_{\rm C1s}$ for the CO molecule, we solve  Equation~\eqref{eq:qpequationdeltaH} with $\Delta \rm H_{\rm C1s}$, whereas we solve it with  $\Delta \rm H_{\rm O1s}$ for $\epsilon^{\rm QP}_{\rm O1s}$. In the case of \ce{HCOOH}, we determine $\Delta \rm H_{\rm core}$ for each O separately. 

In a $G_{\Delta \rm H}W_0$ calculation, we calculate $\Delta \rm H$ once, insert it in Equation~\eqref{eq:qpequationdeltaH} and iterate the latter as in a regular $G_0W_0$ calculation. The shift $\Delta \rm H$ is kept constant during the iteration of the QP equation. Compared to a $G_0W_0$ calculation, the computation of $\Delta \rm H$ is the only computational overhead that we introduced. The  computational cost of a $G_{\Delta \rm H}W_0$ calculation is thus practically the same as for a $G_0W_0$ calculation. 
The $G_{\Delta \rm H}W_0$ method can be viewed as a one-diagonal element correction in the context of the ev$GW_0$ method.

\begin{figure*}[t!]
    \centering
    \includegraphics[width=0.99\textwidth]{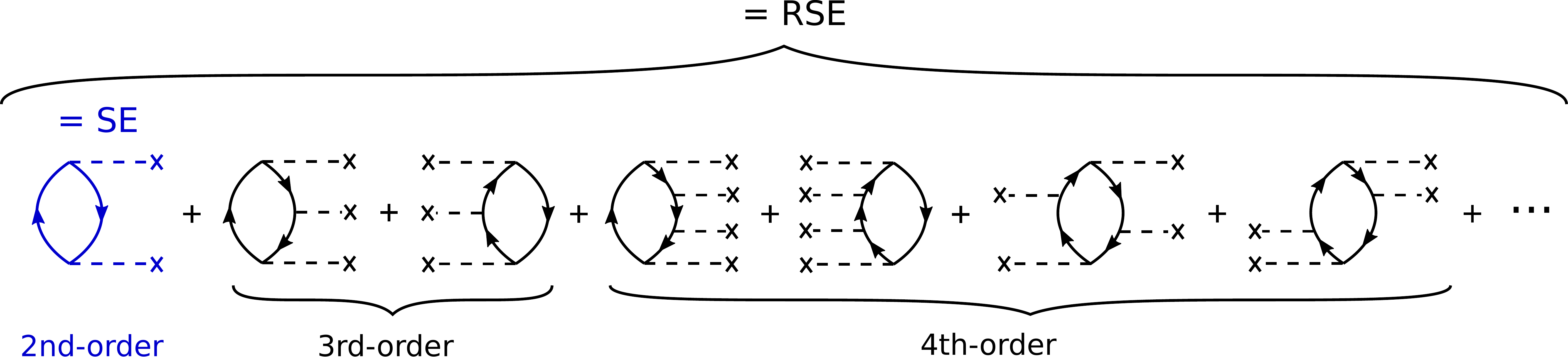}
    \caption{Goldstone diagrammtic for the renormalized singles contributions. Dashed lines, which end with a cross denote the matrix elements $\braket{\psi_p|v^{\text{HF}}-v^{\text{Hxc}}|\psi_q}$}
    \label{fig:diagrams}
\end{figure*}

\subsection{Renormalized singles excitations in RPA}

The RS Green's function approach is based on the same idea as the RSE corrections in RPA. Since  we consider the introduction of the RS concept more illustrative for total than QP energies, we will briefly summarize the key equations derived by Ren \textit{et al.}\cite{renRandomPhaseApproximationElectron2011,renRandomphaseApproximationIts2012,renRenormalizedSecondorderPerturbation2013} in the context of Rayleigh-Schr\"odinger perturbation theory (RSPT), before proceeding with the RS Green's function approaches in Section~\ref{subsec:theory:rsgw}.\par 

In RSPT, the interacting $N$-electron Hamiltonian $\hat{H}$ is partitioned into a non-interacting mean-field Hamiltonian $\hat{H}_0$ and an interacting perturbation $\hat{H}'$.
\begin{align}
  \hat{H}_0 & = \sum_i^N\hat{h}^0(j) = \sum_{j=1}^N\left[-\frac{1}{2}\nabla_j^2 +v_{\text{ext}}(\mathbf{r}_j)+v_j^{\text{Hxc}}\right]\label{eq:singlepartHam}\\
  \hat{H}' & = \sum^N_{j<k}\frac{1}{|\mathbf{r}_j -\mathbf{r}_k| }-\sum_{j=1}^N v_j^{\text{Hxc}}
\end{align}
where $\hat{h}_0$ is the single-particle Hamiltonian of the mean field reference. The Hamiltonian $\hat{h}_0$  includes the kinetic term, the external potential $v_{\text{ext}}$ and the mean-field potential $v^{\text{Hxc}}$. The latter can be the single-particle potential from HF or KS-DFT and contains the Hartree and exchange-correlation terms.
Following standard perturbation theory, single-excitations (SE) contribute to the second-order correlation energy and are given by\cite{renRandomPhaseApproximationElectron2011,renRenormalizedSecondorderPerturbation2013} 
\begin{align}
 E_c^{\text{SE}} &= \sum_i^{\text{occ}}\sum_a^{\text{virt}} \frac{|\braket{\Psi_0|\hat{H}'|\Psi_i^a}|^2}{\epsilon_i^0-\epsilon_a^0}\label{eq:totenergySE1}\\
 & =  \sum_i^{\text{occ}}\sum_a^{\text{virt}} \frac{|\braket{\psi_i^0|v^{\text{HF}}-v^{\text{Hxc}}|\psi_a^0}|^2}{\epsilon_i^0-\epsilon_a^0}\label{eq:totenergySE2}
\end{align}
where $\ket{\Psi_0}$ is the Slater determinant for the ground state configuration and  $\ket{\Psi_i^a}$ for the singly excited configuration. The orbitals $\psi{_{i/a}^0}$ and corresponding orbital energies $\epsilon_{i/a}^0$ are the ones of the $\hat{h}^0(j)$ operator. The derivation of Equation~\eqref{eq:totenergySE2} from Equation~\eqref{eq:totenergySE1} is given in detail in the supporting information (SI) of Ref.~\citenum{renRandomPhaseApproximationElectron2011}. As evident from Equation~\eqref{eq:totenergySE2}, the singles correction vanishes if $v^{\text{Hxc}}$ is the HF mean-field potential, which is a consequence of Brillouin's theorem.\cite{szaboModernQuantumChemistry2012}

The energy $E_c^{\text{SE}}$ is only the second-order correction to the correlation energy, as shown in Figure~\ref{fig:diagrams}. The infinite summation of the higher-order diagrams yields the renormalized singles excitation (RSE) correction. The derivation of the RSE correlation energy is given in detail in Ref.~\citenum{renRenormalizedSecondorderPerturbation2013}. To summarize briefly the procedure, the Fock matrix is evaluated with the KS orbitals. Subsequently, the occupied and unoccupied blocks of this matrix  are diagonalized separately (subspace diagonalization), which yields a new set of (RS) eigenvalues and orbitals. Replacing $\psi_{i/a}$ and $\epsilon_{i/a}$ in Equation~\ref{eq:totenergySE2} with the RS eigenvalues and orbitals yields the RSE correlation energy. 

\subsection{\label{subsec:theory:rsgw}RS Green$'$s function $\boldsymbol{GW}$ approaches}
In analogy to the RPA RSE correction, 
the RS Green's function $G_{\text{RS}}$ is designed as an effective non-interacting reference system that includes all the single contributions. 
The RS Green's function is defined as
\begin{equation}
    G_{\text{RS}}^{-1} = G_0^{-1} - P(\Sigma_{\text{HF}}[G_{0}]-v^{\text{Hxc}})P - Q(\Sigma_{\text{HF}}[G_{0}]-v^{\text{Hxc}})Q \text{,}
    \label{eq:GRS}
\end{equation}
where $P=\sum_{i}^{\text{occ}}|\psi^{0}_{i}\rangle\langle\psi^{0}_{i}|$ is the projection into the occupied orbital space and $Q=I-P$ is the projection into the virtual orbital space.
$\Sigma_{\text{HF}}[G_{0}]$ refers to a HF-like self-energy constructed with $G_0$, which is usually the KS Green's function. 
$\Sigma_{\text{HF}}$ is the sum of the Hartree self-energy $ \Sigma^{\text{H}}$ and the exchange self-energy $\Sigma^x$, i.e., $\Sigma_{\text{HF}}=\Sigma^{\text{H}} + \Sigma^x$, 
where $\Sigma^{\text{H}}_{mn}=\sum_i^{\text{occ}} \Braket{\psi^0_m\psi^0_i|\psi^0_n\psi^0_i}$ and $\Sigma^x_{mn}=-\sum_i \Braket{\psi^0_m\psi^0_i|\psi^0_i\psi^0_n}$. Note that both are built with the mean-field orbitals $\psi_n^0$ provided by, e.g., KS-DFT. $v^{\text{Hxc}}$ is the single-particle Hartree-exchange-correlation potential defined in Equation~\eqref{eq:singlepartHam}. If the potential $v^{\text{Hxc}}$ is the one from HF and if $G_0$ is the HF Green's function, then the second and third term on the right-hand-side of Equation~\eqref{eq:GRS} vanish and $G_{\text{RS}}$ corresponds to the HF Green's function, which is again a consequence of Brillouin's theorem. $G_{\text{RS}}$ includes the singles contributions, which are one source of the starting point dependence. As we showed previously,\cite{jinRenormalizedSinglesGreen2019} the dependence on the DFA is therefore reduced in the $G_{\text{RS}}W_0$ scheme.

The RS Green's function is given as the solution of the two projected equations in the occupied orbital subspace\cite{jinRenormalizedSinglesGreen2019}
\begin{equation}
P(G_{\text{RS}}^{-1})P=P(G_{0}^{-1})P+P(\Sigma_{\text{HF}}[G_{0}]-v^{\text{Hxc}})P\text{,}
\end{equation}
and the virtual orbital subspace
\begin{equation}
Q(G_{\text{RS}}^{-1})Q=Q(G_{0}^{-1})Q+Q(\Sigma_{\text{HF}}[G_{0}]-v^{\text{Hxc}})Q\text{.}
\end{equation}

In practice, $G_{\text{RS}}$ is obtained by a similar subspace diagonalization procedure as used for the RSE total energy corrections. The KS density matrix is used to construct the HF Hamiltonian $\hat{H}_{\text{HF}}$,
which defines the RS Hamiltonian $\hat{H}_{\text{RS}}=\hat{H}_{\text{HF}}[G_{0}]$. 
The equations for the occupied
\begin{equation}
    P(\hat{H}_{\text{HF}}[G_{0}])P|\psi_{i}^{\text{RS}}\rangle =
    \epsilon_{i}^{\text{RS}}P|\psi_{i}^{\text{RS}}\rangle \text{,}\label{eq:rs_occ}
\end{equation}
and virtual subspace
\begin{equation}
    Q(\hat{H}_{\text{HF}}[G_{0}])Q|\psi_{a}^{\text{RS}}\rangle =
    \epsilon_{a}^{\text{RS}}Q|\psi_{a}^{\text{RS}}\rangle \label{eq:rs_vir}
\end{equation}
are diagonalized separately.\cite{jinRenormalizedSinglesGreen2019} The subspace diagonalization yields the RS eigenvalues $\epsilon^{\text{RS}}_n$ and corresponding eigenvectors $\psi^{\text{RS}}_n$ and is performed only once. 
The RS Green's function is computed with the RS eigenvalues and orbitals, 
\begin{equation}
 G_{\text{RS}}(\mathbf{r},\mathbf{r}',\omega) =
 \sum_m \frac{\psi_{m}^{\text{RS}}(\mathbf{r}) \psi_{m}^{\text{RS}}(\mathbf{r'})}{\omega - \epsilon_{m}^{\text{RS}} - i\eta \mathrm{sgn}(\epsilon_{\mathrm{F}} - \epsilon_{m}^{\text{RS}})} \text{,}
\end{equation}
and is diagonal in the occupied subspace and the virtual subspace.

In the $G_{\text{RS}}W_0$ approach,\cite{jinRenormalizedSinglesGreen2019}
the RS Green's function is used as a new starting point
and the screened interaction is calculated with the KS
Green's function.
The correlation part of the $G_{\text{RS}}W_0$ self-energy is\cite{jinRenormalizedSinglesGreen2019}
\begin{equation}
        \Sigma^{c,G_{\text{RS}}W_0}_{n} (\omega) = 
        \sum_m \sum_{s} \frac{| \Braket{\psi^0_n\psi_m^0 | \rho^0_s} |^2 }{\omega - \epsilon^{\text{RS}}_{m}
        + (\Omega^0_s - i\eta) \text{sgn} (\epsilon^{\text{RS}}_{m} - \epsilon_{\mathrm{F}})} \text{, }
    \label{eq:grsw0_se}
\end{equation}
where $\rho^0_s$ and $\Omega^0_s$ are the transition densities and the RPA excitation energies calculated with the KS Green's function. 
Then the QP equation for $G_{\text{RS}}W_0$ is
\begin{equation}\label{eq:qpe_grsw0}
    \epsilon^{\text{QP}}_{n}=\epsilon_{n}^{0} + \mathrm{Re} 
    \braket{ \psi_{n}^0 |
    \Sigma^{G_{\text{RS}}W_0} (\epsilon_{n}^{\text{QP}})- v^{\mathrm{xc}}
     | \psi_{n}^0 } \text{.}
\end{equation}
Note that Equations~\eqref{eq:grsw0_se} and \eqref{eq:qpe_grsw0} are slightly different from the original QP equation for $G_{\text{RS}}W_0$,\cite{jinRenormalizedSinglesGreen2019} where we used the RS eigenvectors $\psi_n^{\text{RS}}$. Here we use for simplicity the KS instead of the RS orbitals because we found that the difference induced upon using RS orbitals is marginal.\cite{jinRenormalizedSinglesGreen2019} Since we use the KS orbitals, the exchange part of the $G_{\text{RS}}W_0$ self-energy is the same as in $G_0W_0$, see Equation~\eqref{eq:sigmax}.

In this work we introduce a new approach that uses the RS Green's function as a new starting point and also calculates the screened Coulomb interaction with the RS Green's function, 
denoted as $G_{\text{RS}}W_{\text{RS}}$.
This means that $W_{\text{RS}}$ is obtained by inserting the RS Green's function into the RPA equation.
Similar to $G_{\text{RS}}W_0$, 
the exchange part of the $G_{\text{RS}}W_{\text{RS}}$ self-energy is also the same as $G_0W_0$, but the correlation part of the $G_{\text{RS}}W_{\text{RS}}$ self-energy becomes
\begin{equation}
        \Sigma^{c,G_{\text{RS}}W_{\text{RS}}}_{n} (\omega) = 
        \sum_m \sum_{s} \frac{|\Braket{\psi^0_n \psi^0_m | \rho^{\text{RS}}_s}| ^2 }{\omega - \epsilon^{\text{RS}}_{m}
        - (\Omega^{\text{RS}}_s - i\eta) \text{sgn} (\epsilon^{\text{RS}}_{m} - \epsilon_{\mathrm{F}})} \text{, }
    \label{eq:grswrs_se}
\end{equation}
where $\rho^{\text{RS}}_s$ and $\Omega^{\text{RS}}_s$ are the transition densities
and the RPA excitation energies calculated with the RS Green's function.
Then the QP equation for the $G_{\text{RS}}W_{\text{RS}}$ approach follows as
\begin{equation}\label{eq:qpe_grswrs}
    \epsilon^{\text{QP}}_{n}=\epsilon_{n}^{0} + \mathrm{Re} 
    \braket{ \psi_{n}^0 |
    \Sigma^{G_{\text{RS}}W_{\text{RS}}} (\epsilon_{n}^{\text{QP}})- v^{\mathrm{xc}}
     | \psi_{n}^0 } \text{.}
\end{equation}
The construction of the RS Green's function scales only as $\mathcal{O}(N^3)$ with respect to system size $N$. The overall scaling of $G_{\text{RS}}W_{\text{RS}}$ depends on the frequency integration technique, which is $O(N^6)$ for a fully analytic evaluation of the self-energy using Equation~\eqref{eq:g0w0sigma_pole}, but $O(N^4)$ (valence states) or $O(N^5)$ (core states) with the contour deformation approach.\cite{govoniLargeScaleGW2015,golzeCoreLevelBindingEnergies2018,zhuAllElectronGaussianBasedG0W02021} Given that the same frequency integration is used for $G_0W_0$ and $G_{\text{RS}}W_{\text{RS}}$, the computational cost of a $G_{\text{RS}}W_{\text{RS}}$ calculation is only marginally larger than for a $G_0W_0$ calculation.

%

\section{COMPUTATIONAL DETAILS}\label{sec:computation}
Core-level calculations were performed at the $G_0W_0$, ev$GW_0$, ev$GW$, $G_{\Delta \rm H}W_0$ and $G_{\rm RS}W_{\rm RS}$ level for the CORE65 benchmark set,\cite{golzeAccurateAbsoluteRelative2020} which contains 65 1s excitations energies of 32 small molecules (30$\times$C1s, 21$\times$O1s, 11$\times$N1s and 3$\times$F1s). Geometries and experimental reference values were taken from Ref.~\citenum{golzeAccurateAbsoluteRelative2020}. Additionally, we studied the ethyl trifluoroacetate molecule. The structure of the latter was obtained upon request from the authors of Ref.~\citenum{kleinNutsBoltsCorehole2021} and is available in the SI. 

The $G_0W_0$, ev$GW_0$, ev$GW$ and $G_{\Delta \rm H}W_0$ calculations were carried out with the FHI-aims program package,\cite{blumInitioMolecularSimulations2009,havuEfficientIntegrationAllelectron2009} which is based on numeric, atom-centered orbitals (NAOs). The $G_0W_0$ and ev$GW_0$ data were extracted from our previous work,\cite{golzeAccurateAbsoluteRelative2020} while the ev$GW$ and $G_{\Delta \rm H}W_0$ data were generated for this benchmark study. In the FHI-aims calculations, the contour deformation technique\cite{govoniLargeScaleGW2015,golzeCoreLevelBindingEnergies2018,golzeGWCompendiumPractical2019} is used to evaluate the self-energy, using a modified Gauss-Legendre grid\cite{renResolutionofidentityApproachHartree2012} with 200 grid points for the imaginary frequency integral part. The $G_{\rm RS}W_{\rm RS}$ calculations were performed with the QM4D program.\cite{qm4d} In QM4D, the $GW$ self-energy integral is calculated fully analytically, see Equation~\eqref{eq:g0w0sigma_pole}. In FHI-aims and also in QM4D, the QP equation is always solved iteratively.

For ev$GW_0$, ev$GW$ and $G_{\Delta \rm H}W_0$, we used the Perdew-Burke-Ernzerhof (PBE)\cite{perdewGeneralizedGradientApproximation1996} functional for the underlying DFT calculation, while the $G_0W_0$ calculations employ the PBEh($\alpha$) hybrid functional\cite{atallaHybridDensityFunctional2013} with  45\% exact exchange ($\alpha=0.45$). The $G_{\rm RS}W_{\rm  RS}$ calculations were performed with PBE and three different hybrid functionals, namely PBE0,\cite{adamoReliableDensityFunctional1999,ernzerhofAssessmentPerdewBurke1999} PBEh($\alpha=0.45)$ and B3LYP.\cite{beckeDensityFunctionalThermochemistry1993,leeDevelopmentColleSalvettiCorrelationenergy1988} Note that PBE0 corresponds to PBEh($\alpha=0.25$).

All $GW$ results are extrapolated to the complete basis set limit using the Dunning basis set family cc-pV$n$Z.\cite{dunningGaussianBasisSets1989,wilsonGaussianBasisSets1996} Following Ref.~\citenum{golzeAccurateAbsoluteRelative2020}, the extrapolation is performed by a linear regression with respect to the inverse of the total number of basis functions. A four-point extrapolation with $n=3-6$ is performed for $G_0W_0$, ev$GW_0$, ev$GW$ and $G_{\Delta \rm H}W_0$.  For $G_{\rm RS}W_{\rm RS}$,  we use only two points ($n=3,4$) due to computational limitations. We verified that this two-point extrapolation deviates only by 0.1~eV from the four-point scheme on average. The cc-pV$n$Z family are contracted Gaussian-type orbitals (GTOs), which can be considered as a special case of an NAO and are treated numerically in FHI-aims. Note that the cc-pV$n$Z basis sets are treated as spherical GTOs in FHI-aims, whereas in QM4D they are processed as pure Cartesian GTOs. Both codes use the resolution-of-the-identity (RI) approach with the Coulomb metric (RI-V).\cite{vahtrasIntegralApproximationsLCAOSCF1993} In FHI-aims, the RI auxiliary basis sets are generated on-the-fly as described in Ref.~\citenum{renResolutionofidentityApproachHartree2012}. For the QM4D calculations, the corresponding RI basis sets for cc-pVTZ and cc-pVQZ from Ref.~\citenum{weigendEfficientUseCorrelation2002} were used.

Relativistic effects were included for all calculations as post-processing step following the approach in Refs.~\citenum{golzeAccurateAbsoluteRelative2020} and \citenum{kellerRelativisticCorrectionScheme2020}, i.e., we performed a non-relativistic $GW$ calculation on top of a non-relativistic KS-DFT calculation and added the corrective term derived in Ref.~\citenum{kellerRelativisticCorrectionScheme2020} to the $GW$ QP energies. The magnitude of the corrections increase with the atomic number and ranges from 0.12~eV for C1s to 0.71~eV for F1s. The relativistic corrections were derived for a free neutral atom at the PBE level and were obtained by evaluating the difference between the 1s eigenvalues from the radial KS and the 4-component Dirac-KS equation.

\section{RESULTS AND DISCUSSIONS}\label{sec:result}

\subsection{Solution behavior}\label{subsec:graphical_solutions}

\begin{figure*}[t]
    \centering
    \includegraphics[width=0.99\textwidth]{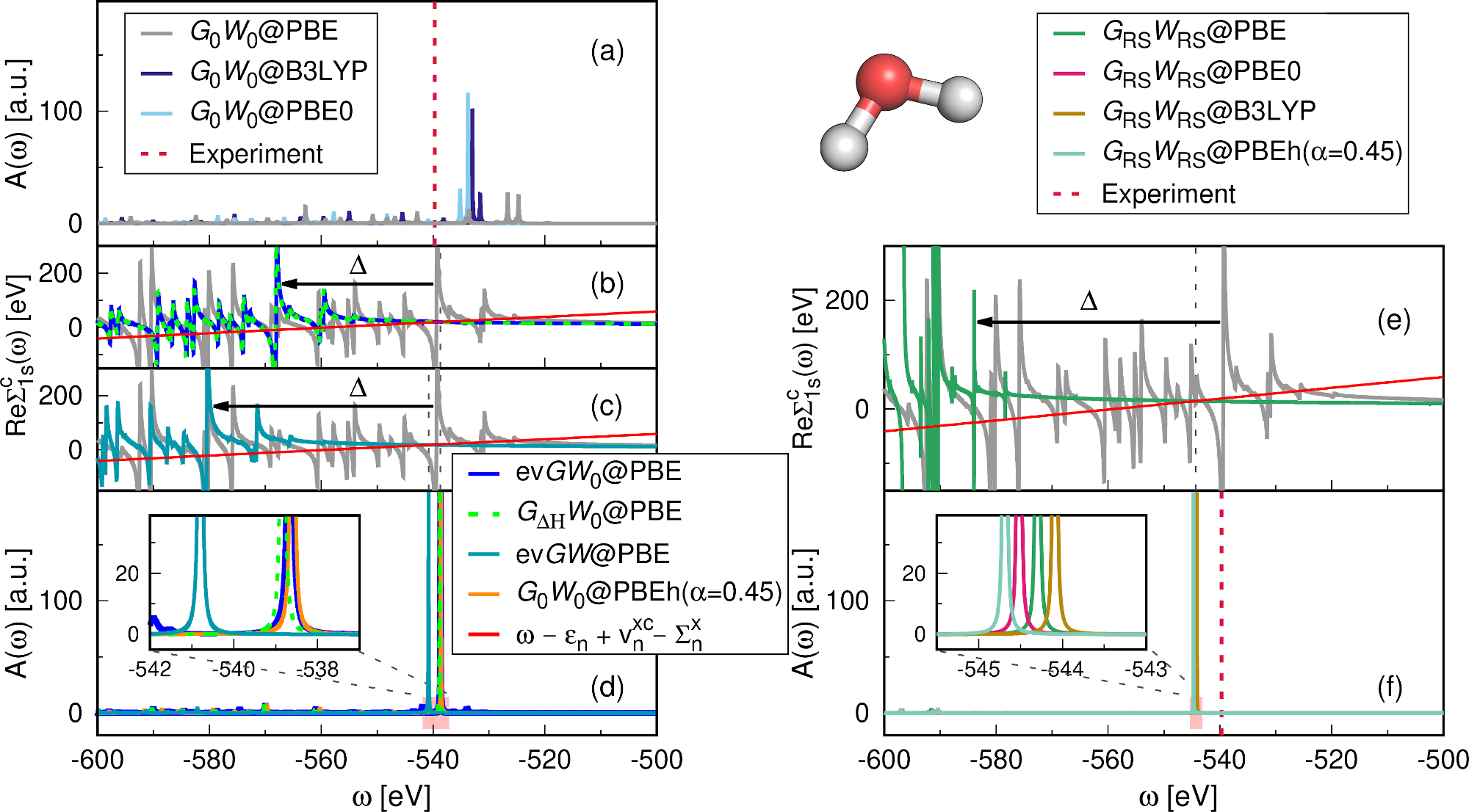}
    \caption{O1s excitation for a single water molecule from $G_0W_0$, ev$GW_0$, $G_{\Delta \text{H}}W_0$, ev$GW$
  and $G_{\mathrm{RS}}W_{\mathrm{RS}}$ computed at the cc-pVQZ level. (a) Spectral function $A(\omega)$ (Equation~\eqref{eq:Aomegatrace}) from $G_0W_0$ using starting points with no exact exchange or a low amount. (b,c) Real part of the correlation self-energy $\Sigma^c(\omega)$ in ev$GW_0@$PBE, $G_{\Delta \text{H}}W_0$ and ev$GW@$PBE. Diagonal matrix elements $\text{Re}\Sigma^c_n(\omega)=\braket{\psi_n|\text{Re}\Sigma^c(\omega)|\psi_n}$ for the oxygen 1s orbital. The intersection with the red line is the graphical solution of Equation~\eqref{eq:qpequation}. The vertical gray-dashed line indicates the QP solution of this graphical solution and $\Delta$ indicates the shift with respect to $G_0W_0@$PBE (gray) (d) Spectral functions in  ev$GW_0@$PBE, $G_{\Delta \text{H}}W_0@$PBE, ev$GW@$PBE and $G_0W_0@$PBEh($\alpha$=0.45). (e) Self-energy matrix elements $\text{Re}\Sigma^c_n(\omega)$ in $G_{\mathrm{RS}}W_{\mathrm{RS}}@$PBE (green) compared to $G_0W_0@$PBE (gray). (f) Spectral functions $A(\omega)$ in $G_{\mathrm{RS}}W_{\mathrm{RS}}$ with different starting points.}
    \label{fig:solutionbehavior}
\end{figure*}

In our previous work~\cite{golzeAccurateAbsoluteRelative2020} we showed that standard $G_0W_0$ calculations starting from a GGA functional, which are routinely applied to valence states, lead to an erroneous multi-solution behavior for deep core states. It is thus important to confirm that the respective $GW$ flavors yield indeed a unique solution. Only looking at the QP energies obtained by iterating Equation~\eqref{eq:qpequation} is typically not enough to verify the latter. Detailed insight into the solution behavior of $GW$-based methods can be obtained by: i) plotting the real part of the correlation self-energy $\Sigma^c$ and ii) plotting the spectral function $A(\omega)$ as defined in Equation~\eqref{eq:Aomegatrace}. In Figure~\ref{fig:solutionbehavior}, we investigate $A(\omega)$ and the diagonal matrix elements $\Sigma^c_n(\omega)$ for the 1s oxygen orbital of a single water molecule. Results are shown for $G_0W_0$ and $G_{\mathrm{RS}}W_{\mathrm{RS}}$  with different starting points (PBE, PBE0, B3LYP, PBEh($\alpha=0.45$)) as well as partial self-consistent schemes, namely ev$GW_0$, ev$GW$ and $G_{\Delta\mathrm{H}}W_0$, using PBE for the underlying DFT calculation. \par
We start our discussion with the $G_0W_0$ spectral functions and self-energy elements displayed in Figure~\ref{fig:solutionbehavior}(a,b,d), where we reproduced for convenience the $G_0W_0@$PBE, $G_0W_0@$PBE0 and $G_0W_0@$PBEh($\alpha=0.45)$ results, which were also presented in Ref.~\citenum{golzeAccurateAbsoluteRelative2020}. Figure~\ref{fig:solutionbehavior}(b) shows the self-energy from a $G_0W_0@$PBE (gray line), which exhibits many poles. The poles are broadened by the $\eta$-term in Equation~\eqref{eq:G0} and thus appear as spikes in the self-energy. For $G_0W_0@$PBE we find that the poles are located in the frequency region where the QP solution is expected (around 540~eV). As already outlined in Section~\ref{subsec:g0w0}, we can obtain the graphical solution of Equation~\eqref{eq:qpequation} by finding the intersections with the straight line $\omega-\varepsilon_{n}^0+v_{n}^{xc}-\Sigma_{n}^x$. For $G_0W_0$, we observe many intersections, which are all valid solutions of the QP equation. The corresponding spectral function in Figure~\ref{fig:solutionbehavior}(a) shows many peaks with equal spectral weight, but no clear main peak that could be assigned to the QP excitation in contrast to the experiment, where a sharp peak at 539.7~eV~\cite{siegbahnESCAAppliedFree1969} is observed. A main peak starts to emerge for hybrid functional starting points, such as PBE0 ($\alpha=0.25$) and B3LYP ($\alpha=0.20$). However, $G_0W_0@$PBE0 and $G_0W_0@$B3LYP still yield an unphysical second peak, which carries a large fraction of the spectral weight.\par
As already discussed in Ref.~\citenum{golzeAccurateAbsoluteRelative2020}, the reason for this unphysical behavior is the overlap of the satellite spectrum with the QP peak. Satellites are due to multielectron excitations that accompany the photoemission process, e.g., shake-up satellites, which are produced when the core photoelectron
scatters a valence shell electron to a higher unoccupied energy level.\cite{sankariHighResolution1s2006,schirmerHighenergyPhotoelectron1s1987}
These peaks appear as series of smaller peaks at higher energies than the QP energy. For molecules, the spectral weight of these peaks is orders of magnitudes smaller than for the main excitation.\cite{schirmerHighenergyPhotoelectron1s1987} Satellites occur in frequency regions where the real part of the self-energy has poles. As demonstrated in, e.g., Ref.~\citenum{golzeGWCompendiumPractical2019}, the imaginary part of the self-energy exhibits complementary peaks at these frequencies (Kramers-Kronig relation). According to Equation~\eqref{eq:Aomegatrace}, large imaginary parts lead to peaks with small spectral weight, i.e., peaks with satellite character. \par
%
The occurrence of pole features around the QP excitations for deep core states can best be understood by analyzing the denominator of the fully analytic expression of the self-energy given in Equation~\eqref{eq:g0w0sigma_pole}. $\Sigma_n^c(\omega)$ has poles on the real axis for $\eta\rightarrow 0$ at $\epsilon^0_i -\Omega_s$ (occupied states $i$) and $\epsilon^0_a +\Omega_s$ (virtual states $a$).  For occupied states,  the eigenvalues $\epsilon^0_i$ are too large (too positive) and the charge neutral excitations $\Omega_s$ are underestimated at the PBE level. As a result, the poles $\epsilon_i^0 - \Omega_s$ are at too positive frequencies and the satellite thus too close to the QP peak. For virtual states, the same reasoning holds for the poles $\epsilon_a+\Omega_s$, but with reversed sign, i.e., the poles are at too small frequencies. The separation between satellites and QP peak is also too small for valence excitations. However, the problem gets progressively worse further away from the Fermi level since the absolute differences between PBE eigenvalues $\epsilon^0_i$ and the QP excitation increases. We demonstrated this for semi-core states,\cite{wilhelmLowScalingGWBenchmark2021} for which a distinct QP peak is still obtained. However, for deep core states the separation becomes finally so small that the satellites merge with the QP peak.\par
The proper separation between QP excitation and satellites can be restored by using an ev$GW_0$ scheme. The ev$GW_0@$PBE self-energy is shown in Figure~\ref{fig:solutionbehavior}(b) (reproduced from Ref.~\citenum{golzeAccurateAbsoluteRelative2020}). Iterating the eigenvalues in $G$ shifts the on-set of the pole structure too more negative frequencies. The overall pole structure is very similar to $G_0W_0@$PBE, but shifted by a constant value of $\Delta=-28.7$~eV. The $G_{\Delta\mathrm{H}}W_0@$PBE self-energy displayed in Figure~\ref{fig:solutionbehavior}(b) is almost identical to ev$GW_0@$PBE. The shift of the pole structure compared to $G_0W_0@$PBE is with $\Delta=-28.8$~eV only slightly larger than for ev$GW_0$. The rigid $\Delta$-shift of the poles features can be understood as follows: In ev$GW_0$ and $G_{\Delta\mathrm{H}}W_0$, the KS eigenvalue $\epsilon_m^0$ are replaced with $\epsilon_m^0+\epsilon_m$ and $\epsilon_m^0+\Delta\mathrm{H}_{\text{1s}}$, respectively, where $\Delta\epsilon_m$ is the self-consistent $GW$ correction for state $m$ and $\Delta\mathrm{H}_{\text{1s}}$ its non-self-consistent approximation for the O1s state (see Equation~\eqref{eq:Hshiftcore}). The poles are consequently located at $\epsilon_{\mathrm{1s}}+\Delta\epsilon_{\mathrm{1s}}-\Omega_s$ and $\epsilon_{\mathrm{1s}}+\Delta\mathrm{H}_{\text{1s}}-\Omega_s$. Since both corrections, $\Delta\epsilon_{\mathrm{1s}}$ and $\Delta\mathrm{H}_{\text{1s}}$ are negative for PBE starting points, the poles, i.e., satellites move to more negative frequencies and are properly separated from the main excitation. The spectral function now exhibits a distinct QP peak as evidenced by  Figure~\ref{fig:solutionbehavior}(d).\par

As discussed in detail previously,\cite{golzeAccurateAbsoluteRelative2020} the effect of eigenvalue self-consistency can be mimicked in a $G_0W_0$ calculation by using a hybrid functional with a high amount of exact exchange $\alpha$. We showed that increasing $\alpha$ progressively shifts the pole features to more negative frequencies. For $\alpha=0.45$, the ev$GW_0@$PBE self-energy is approximately reproduced and the spectral function shows a distinct peak as displayed in Figure~\ref{fig:solutionbehavior}(d). We note here again that values of $\alpha <0.3$ do not yield a clear main peak\cite{golzeAccurateAbsoluteRelative2020} and thus no unique solution, which is also demonstrated in Figure~\ref{fig:solutionbehavior}(a).

The ev$GW$ approach and the $G_{\mathrm{RS}}W_{\mathrm{RS}}$ schemes lead to a significantly stronger shift of the pole features than ev$GW_0@$PBE or $G_{\Delta\mathrm{H}}W_0@$PBE, as shown in Figure~\ref{fig:solutionbehavior}(c) and (e). The spectral functions displayed in Figure~\ref{fig:solutionbehavior}(d) and (f) confirm that ev$GW$ and the $G_{\mathrm{RS}}W_{\mathrm{RS}}$ yield a distinct peak in the spectrum. The RS eigenvalues of the occupied orbitals are more negative and the ones of the virtual orbitals are more positive than the KS eigenvalues. In addition, RPA evaluated with RS fundamental gaps provides larger excitation energies $\Omega_s$.
In $G_{\mathrm{RS}}W_{\mathrm{RS}}$ the poles at $\epsilon_a+\Omega_s$ are shifted in the positive direction and the poles at $\epsilon_i-\Omega_s$ are shifted in the negative direction. Therefore, satellites from these poles are separated from the main peak. 
For $G_{\mathrm{RS}}W_{\mathrm{RS}}$, a unique solution is obtained for all starting points. As we show in Figure~S1 (see SI), the $G_{\mathrm{RS}}W_0$ approach suffers from a multi-solution behavior in the deep core region and cannot be applied for core-level calculations.
\FloatBarrier

\subsection{CORE65 benchmark}\label{subsec:core65}

\FloatBarrier

\begin{figure*}[t!]
\includegraphics[width=0.99\linewidth]{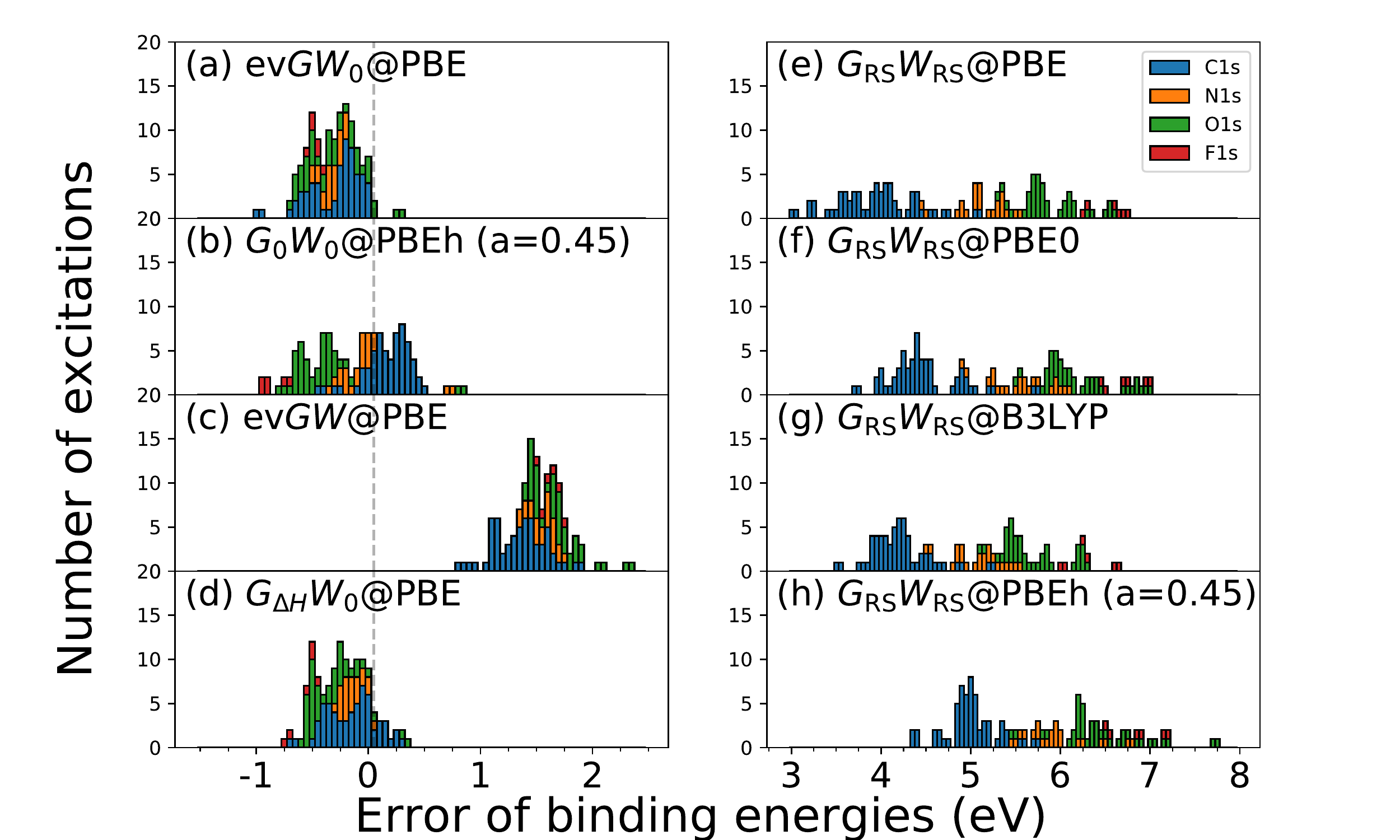}
\caption{Distribution of errors with respect to the experiment for absolute 1s binding energies of the CORE65 benchmark set, where the error is defined as Error$_i$= $\text{CLBE}_i^{\text{theory}}-\text{CLBE}_i^{\text{exp}}$. The histogram is stacked. Eight $GW$ approaches are compared: (a) ev$GW_0$@PBE, (b) $G_0W_0$@PBEh ($\alpha=0.45$), (c) ev$GW$@PBE, (d) $G_{\Delta \text{H}}W_0$@PBE, (e) $G_{\text{RS}}W_{\text{RS}}$@PBE, (f) $G_{\text{RS}}W_{\text{RS}}$@PBE0, (g) $G_{\text{RS}}W_{\text{RS}}$@B3LYP, (h) $G_{\text{RS}}W_{\text{RS}}$@PBEh ($\alpha=0.45$).}
\label{fig:histo_clbe}
\end{figure*}

\begin{figure}
\includegraphics[width=0.9\linewidth]{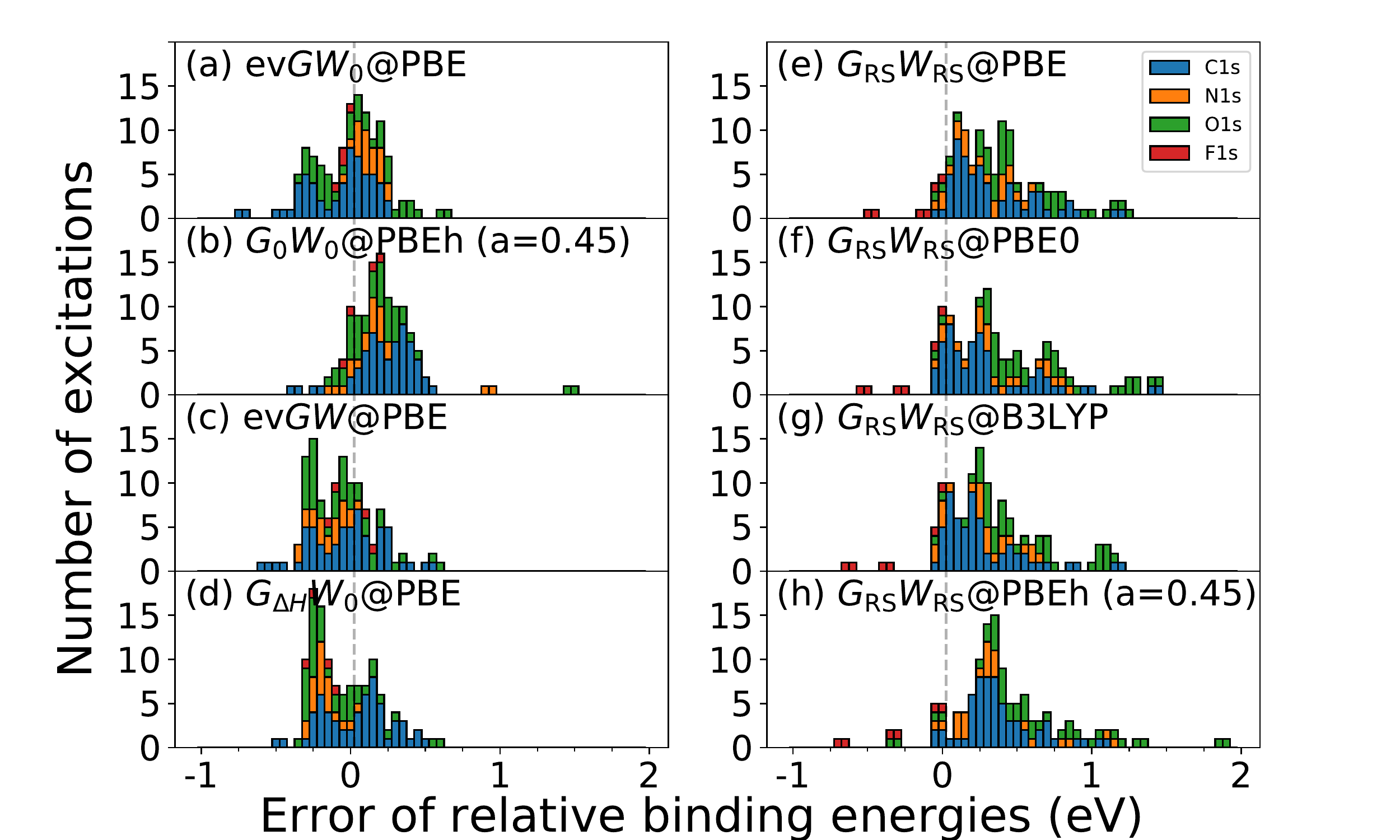}
\caption{Distribution of errors with respect to the experiment for relative 1s binding energies of the CORE65 benchmark set. The histogram is stacked. \ce{CH4}, \ce{NH3}, \ce{H2O} and \ce{CH3F} have been used as reference molecules for C1s, N1s, O1s and F1s respectively. Eight $GW$ approaches are compared: (a) ev$GW_0$@PBE, (b) $G_0W_0$@PBEh($\alpha=0.45$), (c) ev$GW$@PBE, (d) $G_{\Delta \text{H}}W_0$@PBE, (e) $G_{\text{RS}}W_{\text{RS}}$@PBE, (f) $G_{\text{RS}}W_{\text{RS}}$@PBE0, (g) $G_{\text{RS}}W_{\text{RS}}$@B3LYP, (h) $G_{\text{RS}}W_{\text{RS}}$@PBEh($\alpha=0.45$).}
\label{fig:histo_reletive_clbe}
\end{figure}

In the following, we discuss the CORE65 results for the $GW$ schemes for which a physical solution behavior was confirmed in Section~\ref{subsec:graphical_solutions}, namely ev$GW_0@$PBE, ev$GW@$PBE, $G_0W_0@$PBE($\alpha=0.45)$, $G_{\Delta\mathrm{H}}W_0@$PBE and  $G_{\mathrm{RS}}W_{\mathrm{RS}}$ with four different starting points (PBE, PBE0, B3LYP, PBEh($\alpha=0.45)$). The distribution of the errors with respect to experiment are shown in Figure~\ref{fig:histo_clbe} and Figure~\ref{fig:histo_reletive_clbe} for the absolute CLBEs and the relative CLBEs, respectively. The corresponding mean absolute errors (MAE) and the mean errors (ME) are given in Table~\ref{tab:MAE}. The error of excitation $i$ is defined as Error$_i$=$\text{CLBE}_i^{\text{theory}}-\text{CLBE}_i^{\text{experiment}}$.\par
\sloppy Starting the discussion with the absolute CLBEs, we find that ev$GW_0$@PBE, $G_0W_0$@PBEh($\alpha=0.45$) and $G_{\Delta \text{H}}W_0$@PBE yield the best results with error distributions close to zero and MAEs of $\approx 0.3~$eV. The smallest overall MAE of 0.25~eV is obtained with $G_{\Delta \text{H}}W_0$@PBE. Figure~\ref{fig:histo_clbe}(a) and (d) show that the errors from ev$GW_0@$PBE and $G_{\Delta \text{H}}W_0$@PBE are tightly distributed, but mostly negative, i.e., the computed CLBEs are slightly underestimated. Generally, we find that the $G_{\Delta \text{H}}W_0$@PBE scheme reproduces the ev$GW_0@$PBE results almost perfectly. The slight underestimation of the absolute CLBEs by ev$GW_0@$PBE and $G_{\Delta \text{H}}W_0$@PBE might be due to insufficiencies in the cc-pV$n$Z basis sets, which are not captured by the extrapolation procedure. A very recent study\cite{mejia-rodriguezBasisSetSelection2022} with $G_0W_0@$PBEh showed that increasing the number of core functions by, e.g., uncontracting the cc-pV$n$Z basis sets, yields larger absolute CLBEs. The reported increase is in the range of 0.25 to maximal 0.5~eV, indicating that the CLBEs from  ev$GW_0$ and $G_{\Delta \text{H}}W_0$@PBE might be even closer to experiment with core-rich basis sets.\par
The error distribution of $G_0W_0$@PBEh($\alpha=0.45$), which is displayed in Figure~\ref{fig:histo_clbe}(b), is centered around zero, yielding also the smallest overall ME, see Table~\ref{tab:MAE}. Compared to ev$GW_0@$PBE and $G_{\Delta \text{H}}W_0$@PBE, the spread of the $G_0W_0@$PBEh errors is larger and a clustering by species can be observed. The C1s binding energies (BEs) are overestimated, while the N1s, O1s and F1s are increasingly underestimated. This is due to the species dependence of the $\alpha$ parameter, which we discussed in Ref.~\citenum{golzeAccurateAbsoluteRelative2020}. As we showed in Ref.~\citenum{golzeAccurateAbsoluteRelative2020}, including relativistic effects reduces the species dependence of $\alpha$, but does not remove it completely. The optimal $\alpha$ value, $\alpha_\mathrm{opt}$, increases from 0.44 (C1s) to 0.49 (F1s), after including relativistic corrections. For $\alpha<\alpha_{\text{opt}}$, the CLBEs are too small and for $\alpha>\alpha_{\text{opt}}$ too large. As a result,  the C1s BEs are overestimated for $\alpha=0.45$, and O1s and F1s BEs are underestimated.\par
ev$GW$@PBE systematically overestimates the absolute CLBEs by 1-2~eV since iterating also the eigenvalues in $W$ effectively leads to an underscreening. 
At the PBE level, the fundamental gap is underestimated and inserting the PBE eigenvalues in $W$ consequently yields an overscreened potential. However, the overscreening in $W$ compensates the underscreening introduced by the missing vertex corrections. Comparing ev$GW_0@$PBE and ev$GW@$PBE, our observation for deep core-levels agrees with previous work, which found that ev$GW_0$ improves upon $G_0W_0$,\cite{shishkinSelfconsistentGWCalculations2007,byunPracticalGWScheme2019} while ev$GW$ yields too large band gaps~\cite{shishkinSelfconsistentGWCalculations2007} and overly stretched spectra.\cite{maromBenchmarkGWMethods2012} Furthermore, the performance of ev$GW@$PBE for deep core-levels seems to be comparable to higher-level self-consistency schemes such as qs$GW$.\cite{vanschilfgaardeQuasiparticleSelfConsistentGW2006} An exploratory study by Setten \textit{et al.}\cite{vansettenAssessingGWApproaches2018} reported that qs$GW$ overestimates the absolute 1s BEs of small molecules by 2~eV. This indicates that qs$GW$ suffers from similar underscreening effects and that the orbitals inserted in the $GW$ scheme have a minor effect on core-level QP energies.

\begin{table*}[t!]
\small
\setlength\tabcolsep{2.0pt}
\caption{\label{tab:MAE} Mean absolute error (MAE) and mean error (ME) in eV with respect to experiment for absolute and relative CLBEs of the CORE65 benchmark set. The error for excitation $i$ is defined as Error$_i$= $\text{CLBE}_i^{\text{theory}}-\text{CLBE}_i^{\text{exp}}$.  The relative CLBEs are the shifts with respect to a reference molecule, $\Delta\textnormal{CLBE}=\textnormal{CLBE}-\textnormal{CLBE}_{\textnormal{ref\_mol}}$. \ce{CH4}, \ce{NH3}, \ce{H2O} and \ce{CH3F} have been used as reference molecules for C1s, N1s, O1s and F1s respectively.  }
\begin{tabular*}{0.99\linewidth}
{p{0.6cm}ccccccccccccccccc}\toprule

\multirow{2}{*}{\parbox{0.7cm}{\centering core-level}}
&\multicolumn{2}{c}{ev$GW_0@$} 
&\multicolumn{2}{c}{$G_0W_0@$} 
&\multicolumn{2}{c}{ev$GW@$} 
&\multicolumn{2}{c}{$G_{\Delta\mathrm{H}}W_0@$} 
&\multicolumn{8}{c}{$G_{\mathrm{RS}}W_{\mathrm{RS}}@$} \\
\cmidrule{10-17}

&\multicolumn{2}{c}{PBE} 
&\multicolumn{2}{c}{PBEh} 
&\multicolumn{2}{c}{PBE} 
&\multicolumn{2}{c}{PBE} 
&\multicolumn{2}{c}{PBE} 
&\multicolumn{2}{c}{PBE0}
&\multicolumn{2}{c}{B3LYP}
&\multicolumn{2}{c}{PBEh} \\


\cmidrule(l{0.5em}r{0.5em}){2-3} \cmidrule(l{0.5em}r{0.5em}){4-5} \cmidrule(l{0.5em}r{0.5em}){6-7} \cmidrule(l{0.5em}r{0.5em}){8-9} \cmidrule(l{0.5em}r{0.5em}){10-11} \cmidrule(l{0.5em}r{0.5em}){12-13} \cmidrule(l{0.5em}r{0.5em}){14-15} \cmidrule(l{0.5em}r{0.5em}){16-17}

&
\multicolumn{1}{c}{MAE}  & \multicolumn{1}{c}{ME}  &   \multicolumn{1}{c}{MAE}  & \multicolumn{1}{c}{ME} & \multicolumn{1}{c}{MAE}  &\multicolumn{1}{c}{ME} & \multicolumn{1}{c}{MAE}  & \multicolumn{1}{c}{ME} & \multicolumn{1}{c}{MAE}  & \multicolumn{1}{c}{ME} &\multicolumn{1}{c}{MAE}  & \multicolumn{1}{c}{ME} &
\multicolumn{1}{c}{MAE}  & \multicolumn{1}{c}{ME} &
\multicolumn{1}{c}{MAE}  & \multicolumn{1}{c}{ME} &  \\

\midrule

  \multicolumn{17}{c}{Absolute CLBEs}\\
all  & 0.30 &  -0.29  & 0.33 & -0.08 & 1.53 & 1.53 & 0.25   & -0.20 & 4.88 & 4.88  & 5.32 & 5.32 & 4.94 & 4.94 & 5.72 & 5.72 \\
C1s  & 0.27 &  -0.27  & 0.24 & 0.19 & 1.37  & 1.37 & 0.20   & -0.13 & 3.97 & 3.97  & 4.50 & 4.50 & 4.24 & 4.24 & 5.03 & 5.03 \\
N1s  & 0.30 &  -0.30  & 0.16 & -0.01 & 1.58 & 1.58  & 0.14  & -0.13 & 5.13 & 5.13  & 5.59 & 5.59 & 5.12 & 5.12 & 6.00 & 6.00 \\
O1s  & 0.32 &  -0.28  & 0.48 & -0.40 & 1.70 & 1.70 & 0.35   & -0.31 & 5.91 & 5.91  & 6.24 & 6.24 & 5.71 & 5.71 & 7.48 & 7.48 \\
F1s  & 0.44 &  -0.44  & 0.83 & -0.83 & 1.65 & 1.65 & 0.54   & -0.54 & 6.56 & 6.56  & 6.75 & 6.75 & 6.32 & 6.32 & 6.88 & 6.88 \\
 \multicolumn{17}{c}{Relative CLBEs}\\
all  & 0.18   &  0.02 & 0.26    & 0.23 & 0.18 & -0.03  & 0.19 & -0.02  & 0.40 & 0.39  & 0.43 & 0.43  & 0.37 & 0.36  & 0.48 & 0.46    \\
C1s  & 0.18   & -0.05 & 0.29    & 0.25 & 0.19 & -0.01  & 0.20 &  0.07  & 0.36 & 0.36  & 0.33 & 0.33  & 0.30 & 0.30  & 0.41 & 0.41   \\
N1s  & 0.14   &  0.14 & 0.23    & 0.21 & 0.14 & -0.13  & 0.16 & -0.15  & 0.29 & 0.29  & 0.39 & 0.39  & 0.30 & 0.30  & 0.40 & 0.40  \\
O1s  & 0.22   &  0.08 & 0.25    & 0.24 & 0.18 & -0.03  & 0.20 & -0.06  & 0.56 & 0.56  & 0.66 & 0.66  & 0.56 & 0.56  & 0.68 & 0.65    \\
F1s  & 0.05   & -0.05 & 0.11    & 0.10 & 0.11 &  0.02  & 0.16 & -0.16  & 0.16 & -0.10 & 0.16 & 0.00  & 0.20 & 0.02  & 0.22 & 0.00
\\ \bottomrule
\end{tabular*}
\end{table*}

The error distributions for the absolute CLBEs from $G_{\text{RS}}W_{\text{RS}}$ are shown in Figure~\ref{fig:histo_clbe}(e-h) for four different starting points. $G_{\text{RS}}W_{\text{RS}}$ overestimates the absolute CLBEs by 3-8~eV with an MAE between 5-6~eV. 
The reason for the large overestimation is that the RS fundamental gap is too large, which then leads, similarly as in $G_0W_0@$HF,\cite{maromBenchmarkGWMethods2012} to an underscreening in $W$. 
One way to reduce the underscreening is to include corrections for the electron correlation in the RS Hamiltonian, which is dominated by exchange interactions. An alternative strategy is to compensate the underscreening by including vertex corrections, e.g., to use the T-matrix formalism, \cite{martinInteractingElectrons2016,zhangAccurateQuasiparticleSpectra2017,liRenormalizedSinglesGreen2021} where the two-point screened interaction $W$ is replaced with a four-point effective interaction $T$. However, methods such as the T-matrix are computationally much more expensive than $GW$ due to their higher complexity. We recently applied the $G_{\text{RS}}T_{\text{RS}}$ scheme to the CORE65 benchmark set.\cite{liRenormalizedSinglesGreen2021} Comparing $G_{\text{RS}}W_{\text{RS}}$ and $G_{\text{RS}}T_{\text{RS}}$, the overestimation is indeed reduced by $G_{\text{RS}}T_{\text{RS}}$, which yields an ME of $\approx 1.5$~eV. However, the errors for the absolute CLBEs are still an order of magnitude larger than for computationally cheaper schemes such as ev$GW_0$ or $G_{\Delta \text{H}}W_0$, which rely on a very fortunate error cancellation effect that leads to a balanced screening.\par

Furthermore, we find that the overestimation with $G_{\text{RS}}W_{\text{RS}}$ increases with the atomic number, i.e., from the C1s to the F1s excitations. This species dependence is inherited from the KS-DFT calculation.
As shown in Table~S8 in the SI, the deviations of the CLBEs obtained from KS-DFT eigenvalues ($\text{CLBE}_n=-\epsilon_n^0$) generally increase from C1s to F1s for all DFAs. We expect that, e.g., adding correlation contributions to the RS Hamiltonian would also reduces this undesired species dependence.\par

The motivation of the RS approach is the reduction of the starting point dependence. As evident from Figure~\ref{fig:histo_clbe}(e-h), the large overestimation is observed for all DFAs. Based on the MAEs and MEs in Table~\ref{tab:MAE}, it can be shown that the starting point dependence  is on average $< 1$~eV for $\alpha=0.0-0.45$. As shown in Figure~S2 (SI) for a single water molecule, ev$GW_0$ seems to reduce the starting point dependence less. For the same $\alpha$ range, the CLBE changes by $\approx 2$~eV. The direct comparison with $G_0W_0$ is difficult because a unique solution is only obtained for $\alpha > 0.3$. However, we can study the change of the CLBEs for $\alpha=0.3-1.0$, see Figure~S2 (SI), which shows that the starting point dependence is 10~eV with $G_0W_0$ compared to 2.8~eV with ev$GW_0$.
 
\par

Moving now to the relative CLBEs, we observe that ev$GW_0@$PBE, $G_0W_0@$PBEh, ev$GW@$PBE and $G_{\Delta\mathrm{H}}W_0@$PBE yield MAEs of $\approx 0.2-0.3$~eV, and $G_{\mathrm{RS}}W_{\mathrm{RS}}$  MAEs of $0.4-0.5$~eV, see Table~\ref{tab:MAE}. The errors of the relative CLBEs are centered and tightly distributed around zero for ev$GW_0@$PBE, ev$GW@$PBE  and $G_{\Delta\mathrm{H}}W_0@$PBE. The perturbative schemes $G_0W_0@$PBEh and $G_{\mathrm{RS}}W_{\mathrm{RS}}$ slightly overestimate the relative CLBEs. The latter is evident from the positive MEs and the error distributions in Figure~\ref{fig:histo_clbe}(b,e-h), which are not centered at zero, but exhibit a small offset towards positive values. The RS results show a larger spread compared to $G_0W_0@$PBEh and the self-consistent schemes. Furthermore, outliers with errors $>1$~eV are observed for $G_0W_0@$PBEh and in particular for $G_{\mathrm{RS}}W_{\mathrm{RS}}$. The largest outliers are primarily O1s excitations, which originate from the underlying DFT calculation, as evident from Table~S8 (see SI), which shows the MAEs of the relative CLBEs from the KS-DFT eigenvalues. 
For all four functionals (PBE, PBE0, B3LYP, PBEh), we obtained the largest MAE at the KS-DFT level for the O1s excitations. These errors are inherited in the one-shot $G_0W_0$ and $G_{\text{RS}}W_{\text{RS}}$ approaches because of their perturbative nature. 

The chemical shifts between CLBEs of the same atomic type can be smaller than 0.5~eV for second row elements\cite{siegbahnESCAAppliedFree1969} and even as small as 0.1~eV for C1s.\cite{pireauxCoreelectronRelaxationEnergies1976} Therefore, the errors for absolute CLBEs from ev$GW$ and $G_{\mathrm{RS}}W_{\mathrm{RS}}$ are too large to align or resolve experimental XPS spectra, for which reference data are not available. The most promising methods are ev$GW_0@$PBE, $G_0W_0@$PBEh and $G_{\Delta\mathrm{H}}W_0@$PBE. With MAEs between $0.2-0.3$~eV for absolute and relative CLBEs, the accuracy is well within the chemical resolution required to interpret most XPS data.
The disadvantage of the $G_0W_0@$PBEh($\alpha)$ scheme is the need for tuning the $\alpha$ parameter. In addition, the species dependence of $\alpha_{\text{opt}}$ cannot be completely removed. Conversely, the accuracy of ev$GW_0@$PBE and $G_{\Delta\mathrm{H}}W_0@$PBE is species independent. In addition, the already very good agreement of ev$GW_0$@PBE and $G_{\Delta\mathrm{H}}W_0@$PBE with experiment might further improve with core-rich basis sets, as already mentioned before.

Comparing ev$GW_0$, $G_0W_0@$PBEh and $G_{\Delta\mathrm{H}}W_0$, ev$GW_0$ is the computationally most expensive approach. In ev$GW_0$, the eigenvalues are iterated in $G$ (outer loop) and in each step of the outer loop we iterate Equation~\eqref{eq:qpequation} (the QP equation) not only for the core state of interest, but for all states. Using exact methods such as contour deformation, the self-energy needs to be re-evaluated at each iteration step of the QP equation, which usually converges within 10 steps. This implies that even for small molecules we evaluate the self-energy in the ev$GW_0$ procedure several hundred times. The $G_0W_0@$PBEh and $G_{\Delta\mathrm{H}}W_0@$PBE schemes are computationally significantly less expensive. For $G_0W_0$, the QP equation is iterated only once for the core state of interest. Given that the QP equation converges within 10 steps, the total number of self-energy evaluations amounts to 10. The computational cost of $G_{\Delta\mathrm{H}}W_0@$PBE is only marginally larger than for $G_0W_0$. The Hedin shift $\Delta\text{H}$ (see Equation~\eqref{eq:Hshiftcore}) is computed from $\Sigma_{1s}(\epsilon_{1s}^0)$ once before the iteration of Equation~\eqref{eq:qpequationdeltaH}. Given that the latter converges also in 10 steps, the self-energy needs to be calculated 11 instead of 10 times. 

\FloatBarrier
\subsection{ETFA molecule}
\label{subsec:etfa}

\begin{figure*}
\includegraphics[width=.95\linewidth]{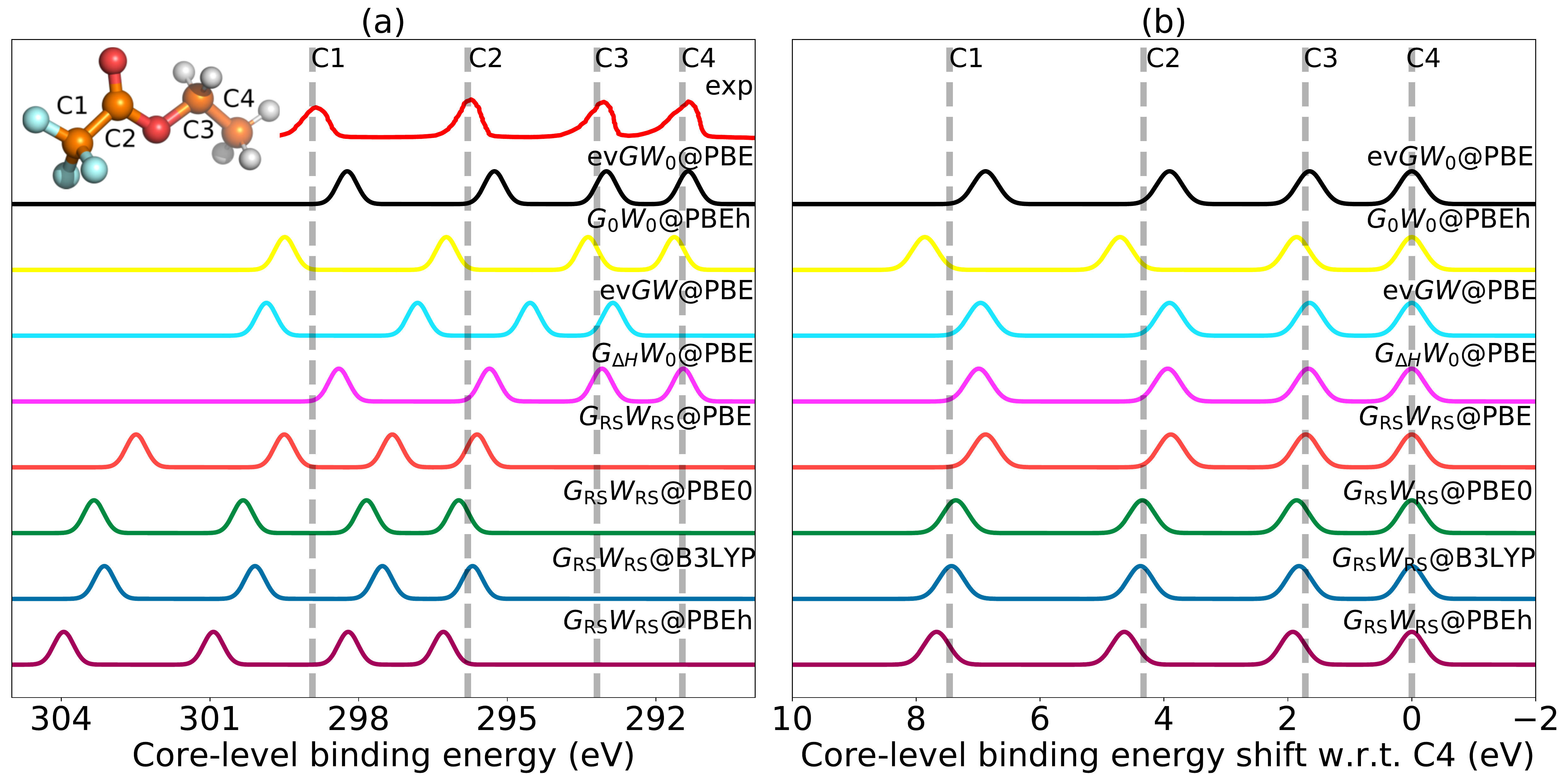}
\caption{\label{fig:etfa}Comparison of the C1s XPS spectrum of ETFA (a) for absolute CLBEs and (b) relative CLBEs obtained from ev$GW_0$@PBE, $G_0W_0$@PBEh($\alpha=0.45$), ev$GW$@PBE, $G_{\Delta \text{H}}W_0$@PBE, $G_{\text{RS}}W_{\text{RS}}$@PBE, $G_{\text{RS}}W_{\text{RS}}$@PBE0, $G_{\text{RS}}W_{\text{RS}}$@B3LYP and $G_{\text{RS}}W_{\text{RS}}$@PBEh($\alpha=0.45$). Dashed lines indicate the experimental reference. }
\end{figure*}

\begin{table*}[t!]
\caption{\label{tab:etfa_abs} Absolute C1s CLBEs for ethyl trifluoroacetate (ETFA), see Figure~\ref{fig:etfa}(a) for the labels. The deviation $\Delta_{\textnormal{exp}}$ with respect to experiment and the corresponding mean (ME) and mean absolute errors (MAE) are computed with the experimental data by Travnikova \textit{et al.},\cite{travnikovaESCAMoleculeHistorical2012} where Error$_i$= $\text{CLBE}_i^{\text{theory}}-\text{CLBE}_i^{\text{exp}}$. }
\begin{tabular*}{0.99\linewidth}{@{\extracolsep{\fill}}lcccccc}\toprule
           & C1  & C2 & C3 & C4 & ME & MAE \\\midrule
experiment (Gelius \textit{et al.}\cite{geliusHighResolutionESCA1973}) & 299.45 & 296.01 & 293.07 & 291.20  & & \\
experiment (Travnikova \textit{et al.}\cite{travnikovaESCAMoleculeHistorical2012})  & 298.93 & 295.80 & 293.19 & 291.47 & & \\
ev$GW_0@$PBE & $-0.70$ & $-0.54$  & $-0.19$ & $-0.12$ & $-0.36$ & 0.36 \\
$G_0W_0@$PBEh($\alpha=0.45$)     &  \phantom{$-$}0.56 & \phantom{$-$}0.54 & \phantom{$-$}0.30 & \phantom{$-$}0.16 & \phantom{$-$}0.39 & 0.39 \\
$G_{\Delta\textnormal{H}}W_0@$PBE & $-0.53$ & $-0.44$ & $-0.10$ & $-0.02$ & $-0.27$ & 0.27\\
ev$GW@$PBE &\phantom{$-$} 0.93 & \phantom{$-$}1.01 & \phantom{$-$}1.35 & \phantom{$-$}1.40 & \phantom{$-$}1.17 & 1.17 \\
$G_{\mathrm{RS}}W_{\mathrm{RS}}@$PBE                 & \phantom{$-$}3.56 & \phantom{$-$}3.70 & \phantom{$-$}4.13 & \phantom{$-$}4.14 & \phantom{$-$}3.88 & 3.88 \\
$G_{\mathrm{RS}}W_{\mathrm{RS}}@$PBE0                & \phantom{$-$}4.41 & \phantom{$-$}4.53 & \phantom{$-$}4.65 & \phantom{$-$}4.51 & \phantom{$-$}4.52 & 4.52 \\
$G_{\mathrm{RS}}W_{\mathrm{RS}}@$B3LYP               & \phantom{$-$}4.20 & \phantom{$-$}4.29 & \phantom{$-$}4.33 & \phantom{$-$}4.23 & \phantom{$-$}4.26 & 4.26 \\
$G_{\mathrm{RS}}W_{\mathrm{RS}}@$PBEh($\alpha=0.45$) & \phantom{$-$}5.02 & \phantom{$-$}5.13 & \phantom{$-$}5.02 & \phantom{$-$}4.82 & \phantom{$-$}5.00 & 5.00 \\
ev$GW_0@$PBE\cite{mejia-rodriguezScalableMolecularGW2021} & $-0.41$ & $-0.18$ & $-0.04$ & $-0.09$ & $-0.18$ & $0.18$ \\
$\Delta$SCAN\cite{kleinNutsBoltsCorehole2021} & $-0.15$ & $-0.08$ & \phantom{$-$}0.05 & \phantom{$-$}0.17 &\phantom{$-$}0.00 & 0.11\\
$\Delta$CCSD(T)\cite{zhengPerformanceDeltaCoupledClusterMethods2019}  & $-0.35$ & $-0.24$ & $-0.31$ & $-0.23$ & $-0.28$ & 0.28
\\ \bottomrule
\end{tabular*}
\end{table*}

\begin{table*}[t!]
\caption{\label{tab:etfa_rel} Relative C1s CLBEs for ethyl trifluoroacetate (ETFA), see Figure.\ref{fig:etfa} for the labels. The deviation $\Delta_{\textnormal{exp}}$ with respect to experiment and the corresponding mean (ME) and mean absolute errors (MAE) are computed with the experimental data by Travnikova \textit{et al.},\cite{travnikovaESCAMoleculeHistorical2012} where $\Delta_{\textnormal{exp}}$= $\Delta\text{CLBE}_i^{\text{theory}}-\Delta\text{CLBE}_i^{\text{exp}}$. The labels of the C atoms are given in the inset of Figure~\ref{fig:etfa}(a).}
\begin{tabular*}{0.99\linewidth}{@{\extracolsep{\fill}}lcccccc}\toprule
           & C1  & C2 & C3 & C4 & ME & MAE \\\midrule
experiment (Gelius \textit{et al.}\cite{geliusHighResolutionESCA1973}) & 8.25 & 4.81 & 1.87 & 0.00\\
experiment (Travnikova \textit{et al.}\cite{travnikovaESCAMoleculeHistorical2012})  & 7.46 & 4.33 & 1.72 & 0.00 \\\midrule
ev$GW_0@$PBE & $-0.58$ & $-0.42$ & $-0.07$ & 0.00 & $-0.36$ & 0.36 \\
$G_0W_0@$PBEh($\alpha=0.45$)     & \phantom{$-$}0.40 & \phantom{$-$}0.38 & \phantom{$-$}0.14 & 0.00 &\phantom{$-$} 0.31 & 0.31  \\
$G_{\Delta\textnormal{H}}W_0@$PBE & $-0.50$ & $-0.42$ & $-0.07$ & 0.00 & $-0.33$ & 0.33\\
ev$GW@$PBE & $-0.47$ & $-0.39$ & $-0.05$ & 0.00 & $-0.30$ & $0.30$ \\
$G_{\mathrm{RS}}W_{\mathrm{RS}}@$PBE                 & $-0.58$ & $-0.44$ & $-0.01$ & 0.00 & $-0.34$ & 0.34 \\
$G_{\mathrm{RS}}W_{\mathrm{RS}}@$PBE0                & $-0.10$ &  \phantom{$-$}0.02 &  \phantom{$-$}0.14 & 0.00 &  \phantom{$-$}0.02 & 0.09 \\
$G_{\mathrm{RS}}W_{\mathrm{RS}}@$B3LYP               & $-0.03$ &  \phantom{$-$}0.05 &  \phantom{$-$}0.10 &0.00 &  \phantom{$-$}0.04 & 0.06 \\
$G_{\mathrm{RS}}W_{\mathrm{RS}}@$PBEh($\alpha=0.45$) &  \phantom{$-$}0.21 &  \phantom{$-$}0.31 &  \phantom{$-$}0.20 & 0.00 &  \phantom{$-$}0.24 & 0.24\\
ev$GW_0@$PBE\cite{mejia-rodriguezScalableMolecularGW2021} & $-0.32$ & $-0.09$ & \phantom{-}0.05 &0.00 & $-0.12$ & 0.15  \\
$\Delta$SCAN\cite{kleinNutsBoltsCorehole2021} & $-0.32$ & $-0.25$ & $-0.12$ & 0.00 & $-0.23$ & 0.23 \\
$\Delta$CCSD(T)\cite{zhengPerformanceDeltaCoupledClusterMethods2019}  & $-0.12$ & $-0.01$ & $-0.08$ & 0.00 & $-0.07$ &0.07
\\ \bottomrule
\end{tabular*}
\end{table*}
We further examine the performance of the eight $GW$ approaches, which we applied to the CORE65 benchmark set in Section~\ref{subsec:core65}, for C1s excitations of the ethyl trifluoroacetate (ETFA) molecule, which is also referred to as the `ESCA molecule' in the literature.\cite{travnikovaESCAMoleculeHistorical2012} The ETFA molecule was  synthesized to demonstrate the potential of XPS for chemical analysis in the late 1960s. It contains four carbon atoms in various chemical environments, see inlet of Figure~\ref{fig:etfa}(a). 
The ETFA molecule presents a challenge because of the extreme variations of the chemical shifts, which range up to $8.0$ \,{eV} and decrease from the \ce{CF3} to the \ce{CH3} end. The four C1s signals are separated by several electronvolts due to the widely different electronegativities of the substituents on the carbon atoms. ETFA is thus an important reference system and was very recently used to benchmark the performance of different functionals in $\Delta$SCF  calculations\cite{kleinNutsBoltsCorehole2021,delesmaChemicalViewXray2018} and $GW$ approaches.\cite{mejia-rodriguezScalableMolecularGW2021} 

In equilibrium, ETFA exists in two dominating conformations (anti-anti and anti-gauche), i.e., each peak in the spectrum is a superposition of the signals from both conformers.\cite{travnikovaESCAMoleculeHistorical2012} However, including the different conformations is primarily important when resolving the vibrational profiles of the peaks, which is not the scope of our benchmarking effort. The experimental conformational shifts are $<0.1$~eV.\cite{travnikovaESCAMoleculeHistorical2012} To ensure direct comparability with the computational data from Ref.~\citenum{kleinNutsBoltsCorehole2021}, we include thus only the anti-anti conformer. \par

The first high-quality experimental spectrum of the free ETFA molecule in gas-phase was reported in 1973,\cite{geliusHighResolutionESCA1973} while new results were published by Travnikova \textit{et al.} in 2012,\cite{travnikovaESCAMoleculeHistorical2012} see Table~\ref{tab:etfa_abs}. Both results were referenced in the most recent studies.\cite{mejia-rodriguezScalableMolecularGW2021,kleinNutsBoltsCorehole2021} The newer results have a higher resolution and are vibrationally resolved. More relevant for this work is that the chemical shifts of the more de-screened carbon atoms are significantly larger smaller than in the older spectrum. The difference between the experimental spectra was attributed to missing correction techniques in early multi-channel plate detectors. We follow here the reasoning of Ref.~\citenum{mejia-rodriguezScalableMolecularGW2021}, pointing out that coupled-cluster results\cite{zhengPerformanceDeltaCoupledClusterMethods2019} are significantly closer to the new experimental data (in particular the chemical shifts). We use thus the data by Travnikova \textit{et al.} as experimental reference.     

The comparison between the experimental spectrum and calculated spectra are shown in Figure~\ref{fig:etfa}. The differences to the experimental peak positions are reported in Table~\ref{tab:etfa_abs} and \ref{tab:etfa_rel}. For the absolute CLBEs, the ETFA predictions are inline with the CORE65 benchmark results. ev$GW_0@$PBE and $G_{\Delta\mathrm{H}}W_0@$PBE slightly underestimate the CLBEs, while ev$GW@$PBE and the RS schemes severely overestimate them, see Figure~\ref{fig:etfa}. $G_0W_0@$PBEh($\alpha=0.45)$ also overestimates the C1s energies slightly. As discussed in Section~\ref{subsec:core65}, this is due to the fact that $\alpha_{\text{opt}}$ is slightly smaller than 0.45 for C1s excitation. C1s excitations are consequently slightly overestimated in $G_0W_0@$PBEh. $G_{\Delta\textnormal{H}}W_0@$PBE provides the best accuracy followed by ev$GW_0@$PBE with an MAE of $0.27$~eV and 0.36~eV, respectively, which is consistent with the conclusion for the CORE65 set benchmark. 

Turning to the relative CLBEs, the three self-consistent schemes and the $G_0W_0@$PBEh approach yield MAEs around 0.3~eV, which is only slightly worse than the CORE65 MAEs for C1s. ev$GW_0$, ev$GW$ and $G_{\Delta\mathrm{H}}W_0$ underestimate all shifts, while $G_0W_0@$PBEh overestimates the shifts of the more  "descreened" C atoms (C1 and C2), see Figure~\ref{fig:etfa}(b). $G_{\mathrm{RS}}W_{\mathrm{RS}}$ also provides good shifts  for ETFA. The RS results are again not completely independent on the starting point. With PBE, the C1s shifts are similarly underestimated as for the self-consistent schemes, while they are slightly too large with PBEh. The RS schemes with the conventional hybrid functionals PBE0 and B3LYP yield, with MAEs below 0.1~eV, the best overall result of the 8 investigated schemes. Furthermore, except for $G_{\mathrm{RS}}W_{\mathrm{RS}}$ with hybrid starting points, we find that our predictions worse  progressively with the electronegativity of the substitutents at the C atoms. The relative CLBEs of C1 (\ce{CF3}) and C2 (carbonyl) seem to be the ones that are more difficult to predict, while the predictions of the C3 (\ce{CH2)} shifts are mostly within 0.1~eV of the experimental references. However, this trend can be also found in the deviations between the experimental references, see Table~\ref{tab:etfa_rel}, i.e., deviations between the experiment values increase with the "descreening" of the C atoms.\par

Finally, we compare our results to previously published computational XPS data.\cite{travnikovaESCAMoleculeHistorical2012,mejia-rodriguezScalableMolecularGW2021,kleinNutsBoltsCorehole2021,vandenbosscheEffectsNonlocalExchange2014,delesmaChemicalViewXray2018,travnikovaEnergyDependentRelativeCross2019,zhengPerformanceDeltaCoupledClusterMethods2019} We focus here on the results from i) the same method (ev$GW_0$), ii) one of the most successful functionals for $\Delta$SCF ($\Delta$SCAN) and iii) from a higher-level method (coupled cluster), see Tables~\ref{tab:etfa_abs} and \ref{tab:etfa_rel}. All three literature results include relativistic effects. The previous ev$GW_0@$PBE study\cite{mejia-rodriguezScalableMolecularGW2021} uses the same correction scheme \cite{kellerRelativisticCorrectionScheme2020} as employed in this work. The $\Delta$SCAN results were obtained with the atomic zeroth-order regular approximation (ZORA).\cite{blumInitioMolecularSimulations2009} For the delta coupled cluster doubles results with triples correction ($\Delta$CCSD(T)),\cite{zhengPerformanceDeltaCoupledClusterMethods2019} scalar relativistic effects were taken into account by an exact tow-component theory. The $\Delta$SCAN and coupled cluster results were computed for the anti-anti conformer, whereas the ev$GW_0$ literature data are actually an average of the CLBEs of both conformers.

The absolute ev$GW_0@$PBE CLBEs from Ref.~\citenum{mejia-rodriguezScalableMolecularGW2021} are less underestimated with respect to experiment in comparison to our results. The largest differences are observed for the descreened C1 and C2 atoms. The previously published ev$GW_0$ results\cite{mejia-rodriguezScalableMolecularGW2021} are not extrapolated, but obtained with a basis sets with more core functions. As already discussed for the CORE65 results, substantially increasing the amount of core functions, seems to cure the (small) systematic underestimation of ev$GW_0$ for absolute CLBEs. The predictions upon including more core functions should change on average by 0.25~eV for C atoms,\cite{mejia-rodriguezBasisSetSelection2022} which is close to the (maximal) 0.3~eV difference we observe here. However, similar to our results, the deviation from experiments are larger for C1/C2 than for C3/C4. Since adding the core functions seems to be more relevant for the descreened environments, the chemical shifts improve, too, see Table~\ref{tab:etfa_rel}. Some of the difference must be also attributed to the inclusion of the anti-gauche conformer in Ref.~\citenum{mejia-rodriguezScalableMolecularGW2021}, which has slightly higher C1s BEs.\cite{travnikovaESCAMoleculeHistorical2012}

We note here that Ref.~\citenum{mejia-rodriguezScalableMolecularGW2021} contains also ETFA results with $G_0W_0@$PBEh($\alpha=0.45)$ employing core-rich basis sets. However, the $\alpha$ values was tuned with respect to experiment using an extrapolation scheme with the cc-pV$n$Z basis sets.\cite{golzeAccurateAbsoluteRelative2020} An insufficiency in the basis set description (i.e. here a systematic underestimation) would be partly absorbed in the $\alpha$ value, which is the reason why our $G_0W_0@$PBEh results will agree better with experiment.

It has been recently shown that the SCAN functional yields excellent absolute and relative CLBEs for molecules\cite{kahkAccurateAbsoluteCoreelectron2019} and solids.\cite{kahkCoreElectronBinding2021} We find that this is also true for the ETFA molecule. $\Delta$SCAN yields the best MAE for absolute CLBEs, which is, however, very close to the ev$GW_0@$PBE results\cite{mejia-rodriguezScalableMolecularGW2021} with core-function-rich basis set. For the relative CLBEs, the $\Delta$SCAN is outperformed by partially self-consistent and RS $GW$ approaches as well as coupled cluster.\par

The absolute $\Delta$CCSD(T) core excitations~\cite{zhengPerformanceDeltaCoupledClusterMethods2019} are underestimated by 0.23-0.35~eV. However, these results were obtained at the cc-pVTZ level and are probably not fully converged.\cite{ambroiseProbingBasisSet2021} It is thus difficult to judge the performance of the method for absolute CLBEs. The chemical shifts on the contrary are often less affected by the basis set choice and $\Delta$CCSD(T) yields together with $G_{\mathrm{RS}}W_{\mathrm{RS}}$(@PBE0 or @B3LYP) MAEs $<0.1$~eV. 

\section{Conclusion}\label{sec:conclusion}
We have presented a benchmark study of different $GW$ approaches for the prediction of absolute and relative CLBEs. In addition to the ev$GW_0@$PBE and $G_0W_0@$PBEh($\alpha=0.45$) methods, which were already investigated in Ref.~\citenum{golzeAccurateAbsoluteRelative2020}, we have included ev$GW@$PBE and two new methods, namely $G_{\Delta\mathrm{H}}W_0$ and $G_{\mathrm{RS}}W_{\mathrm{RS}}$, in our study. $G_{\Delta\mathrm{H}}W_0$ is an adaption of the "Hedin shift"\cite{leeTransitionAdiabaticSudden1999,pollehnAssessmentLessGreater1998} to core levels and can be considered as computationally less expensive approximation to ev$GW_0$. 
In the $G_{\mathrm{RS}}W_{\mathrm{RS}}$ approach, 
the RS Green's function is used as a new starting point and, in contrast to our previous work~\cite{jinRenormalizedSinglesGreen2019}, also used to compute the screened Coulomb interaction. The purpose of introducing the RS scheme is to reduce the dependence on the starting point and the method has thus been tested with four different DFAs (PBE, PBE0, B3LYP and PBEh($\alpha=0.45$).

By investigating the self-energy matrix elements and spectral functions, we have confirmed that ev$GW_0@$PBE, $G_0W_0@$PBEh, $G_{\Delta\mathrm{H}}W_0@$PBE, $G_{\mathrm{RS}}W_{\mathrm{RS}}@$PBE, $G_{\mathrm{RS}}W_{\mathrm{RS}}@$PBE0, $G_{\mathrm{RS}}W_{\mathrm{RS}}@$B3LYP and $G_{\mathrm{RS}}W_{\mathrm{RS}}@$PBEh yield a unique solution. $G_0W_0$ schemes starting from a GGA or hybrid DFT calculation with a low amount of exact exchange do not yield a distinct QP solution.  A meaningful physical solution can thus not be obtained with standard approaches such as $G_0W_0@$PBE, $G_0W_0@$PBE0 and $G_0W_0@$B3LYP for CLBEs. 

We have studied the CORE65 benchmark set and the C1s excitations of the ETFA molecule with all 8 approaches, for which a physically reasonable solution behavior was confirmed. For the CORE65 set, ev$GW_0@$PBE, $G_0W_0@$PBEh and $G_{\Delta\mathrm{H}}W_0@$PBE yield with MAEs of 0.30, 0.33  and 0.25~eV, respectively the best results. ev$GW$ and $G_{\mathrm{RS}}W_{\mathrm{RS}}$ overestimate the absolute CLBEs by several electronvolts and are thus not suitable for the prediction of the absolute binding energies. Nevertheless, the RS approaches significantly reduces the starting point dependence as intended. The relative CLBEs are reasonably reproduced with all methods, but in particular with the eigenvalue self-consistent schemes and $G_{\Delta\mathrm{H}}W_0@$PBE (MAEs $<0.2$~eV). The methods exhibit a similar performance for the ETFA molecule, except that the RS approaches with standard hybrid functionals yield here the best chemical shifts.

The $G_0W_0@$PBEh($\alpha$) approach was introduced in our previous work\cite{golzeAccurateAbsoluteRelative2020} as computationally affordable alternative to ev$GW_0$ that can mimic to some extent the effect of eigenvalue self-consistency in $G$. However, the $\alpha$-tuning is methodologically unsatisfying and the optimal $\alpha$ is dependent on the atomic species. We therefore recommend to use the $G_{\Delta\mathrm{H}}W_0@$PBE approach instead, which is in terms of computational cost comparable to $G_0W_0$.  

Finally, we found that ev$GW_0@$PBE and $G_{\Delta\mathrm{H}}W_0@$PBE systematically underestimate the experiment. Our comparison to the ETFA literature results and very recent work\cite{mejia-rodriguezBasisSetSelection2022} suggest that this slight, but systematic underestimation can be cured by very large, core-rich basis sets, which might improve the agreement with experiment even further. Future work will consider this and focus on the development of compact and computationally efficient NAO basis sets for core-level $GW$ calculations.    

\section*{SUPPORTING INFORMATION}
See the Supporting Information for CORE65 benchmark results, solution behavior of $G_{\mathrm{RS}}W_0$, errors of KS-DFT for predicting absolute CLBEs and relative CLBEs, starting point dependence on the tuning parameter in PBEh, geometry of ETFA.

\begin{acknowledgement}
J. L. acknowledges the support from the National
Institute of General Medical Sciences of the National Institutes of
Health under award number R01-GM061870. W.Y. acknowledges the support
from the National Science Foundation (grant no. CHE-1900338). D.G. acknowledges the Emmy Noether Programme of the German Research Foundation under
project number 453275048 and P. R. the support from the European Union’s Horizon 2020 research and innovation program under Grant Agreement No. 951786 (The NOMAD CoE). Computing time from CSC
-- IT Center for Science (Finland) is gratefully acknowledged. We also thank Levi Keller for providing preliminary data for the ETFA molecule. 
\end{acknowledgement}

\section*{Data Availability Statement}
The data are available in the SI and the input and output files of the FHI-aims calculations are available from the NOMAD data base.\cite{nomad_repo}

\bibliography{reference,software}

\providecommand{\latin}[1]{#1}
\makeatletter
\providecommand{\doi}
  {\begingroup\let\do\@makeother\dospecials
  \catcode`\{=1 \catcode`\}=2 \doi@aux}
\providecommand{\doi@aux}[1]{\endgroup\texttt{#1}}
\makeatother
\providecommand*\mcitethebibliography{\thebibliography}
\csname @ifundefined\endcsname{endmcitethebibliography}
  {\let\endmcitethebibliography\endthebibliography}{}
\begin{mcitethebibliography}{103}
\providecommand*\natexlab[1]{#1}
\providecommand*\mciteSetBstSublistMode[1]{}
\providecommand*\mciteSetBstMaxWidthForm[2]{}
\providecommand*\mciteBstWouldAddEndPuncttrue
  {\def\EndOfBibitem{\unskip.}}
\providecommand*\mciteBstWouldAddEndPunctfalse
  {\let\EndOfBibitem\relax}
\providecommand*\mciteSetBstMidEndSepPunct[3]{}
\providecommand*\mciteSetBstSublistLabelBeginEnd[3]{}
\providecommand*\EndOfBibitem{}
\mciteSetBstSublistMode{f}
\mciteSetBstMaxWidthForm{subitem}{(\alph{mcitesubitemcount})}
\mciteSetBstSublistLabelBeginEnd
  {\mcitemaxwidthsubitemform\space}
  {\relax}
  {\relax}

\bibitem[Bagus \latin{et~al.}(2013)Bagus, Ilton, and
  Nelin]{bagusInterpretationXPSSpectra2013}
Bagus,~P.~S.; Ilton,~E.~S.; Nelin,~C.~J. The Interpretation of {{XPS}} Spectra:
  {{Insights}} into Materials Properties. \emph{Surf. Sci. Rep.} \textbf{2013},
  \emph{68}, 273--304\relax
\mciteBstWouldAddEndPuncttrue
\mciteSetBstMidEndSepPunct{\mcitedefaultmidpunct}
{\mcitedefaultendpunct}{\mcitedefaultseppunct}\relax
\EndOfBibitem
\bibitem[H{\"o}fft \latin{et~al.}(2006)H{\"o}fft, Bahr, Himmerlich, Krischok,
  Schaefer, and Kempter]{hofftElectronicStructureSurface2006}
H{\"o}fft,~O.; Bahr,~S.; Himmerlich,~M.; Krischok,~S.; Schaefer,~J.~A.;
  Kempter,~V. Electronic {{Structure}} of the {{Surface}} of the {{Ionic
  Liquid}} [{{EMIM}}][{{Tf2N}}] {{Studied}} by {{Metastable Impact Electron
  Spectroscopy}} ({{MIES}}), {{UPS}}, and {{XPS}}. \emph{Langmuir}
  \textbf{2006}, \emph{22}, 7120--7123\relax
\mciteBstWouldAddEndPuncttrue
\mciteSetBstMidEndSepPunct{\mcitedefaultmidpunct}
{\mcitedefaultendpunct}{\mcitedefaultseppunct}\relax
\EndOfBibitem
\bibitem[{J.~Villar-Garcia} \latin{et~al.}(2011){J.~Villar-Garcia}, F.~Smith,
  W.~Taylor, Qiu, J.~Lovelock, G.~Jones, and
  Licence]{j.villar-garciaChargingIonicLiquid2011}
{J.~Villar-Garcia},~I.; F.~Smith,~E.; W.~Taylor,~A.; Qiu,~F.;
  J.~Lovelock,~K.~R.; G.~Jones,~R.; Licence,~P. Charging of Ionic Liquid
  Surfaces under {{X-ray}} Irradiation: The Measurement of Absolute Binding
  Energies by {{XPS}}. \emph{Phys. Chem. Chem. Phys.} \textbf{2011}, \emph{13},
  2797--2808\relax
\mciteBstWouldAddEndPuncttrue
\mciteSetBstMidEndSepPunct{\mcitedefaultmidpunct}
{\mcitedefaultendpunct}{\mcitedefaultseppunct}\relax
\EndOfBibitem
\bibitem[Siegbahn(1969)]{siegbahnESCAAppliedFree1969}
Siegbahn,~K. \emph{{{ESCA}} Applied to Free Molecules}; {North-Holland
  Publishing}: {Amsterdam; London}, 1969\relax
\mciteBstWouldAddEndPuncttrue
\mciteSetBstMidEndSepPunct{\mcitedefaultmidpunct}
{\mcitedefaultendpunct}{\mcitedefaultseppunct}\relax
\EndOfBibitem
\bibitem[Barr(2020)]{barrModernESCAPrinciples2020}
Barr,~T.~L. \emph{Modern {{ESCA}}: {{The Principles}} and {{Practice}} of
  {{X-Ray Photoelectron Spectroscopy}}}; {CRC Press}: {Boca Raton}, 2020\relax
\mciteBstWouldAddEndPuncttrue
\mciteSetBstMidEndSepPunct{\mcitedefaultmidpunct}
{\mcitedefaultendpunct}{\mcitedefaultseppunct}\relax
\EndOfBibitem
\bibitem[Aarva \latin{et~al.}(2019)Aarva, Deringer, Sainio, Laurila, and
  Caro]{aarvaUnderstandingXraySpectroscopy2019}
Aarva,~A.; Deringer,~V.~L.; Sainio,~S.; Laurila,~T.; Caro,~M.~A. Understanding
  {{X-ray Spectroscopy}} of {{Carbonaceous Materials}} by {{Combining
  Experiments}}, {{Density Functional Theory}}, and {{Machine Learning}}.
  {{Part I}}: {{Fingerprint Spectra}}. \emph{Chem. Mater.} \textbf{2019},
  \emph{31}, 9243--9255\relax
\mciteBstWouldAddEndPuncttrue
\mciteSetBstMidEndSepPunct{\mcitedefaultmidpunct}
{\mcitedefaultendpunct}{\mcitedefaultseppunct}\relax
\EndOfBibitem
\bibitem[Aarva \latin{et~al.}(2019)Aarva, Deringer, Sainio, Laurila, and
  Caro]{aarvaUnderstandingXraySpectroscopy2019a}
Aarva,~A.; Deringer,~V.~L.; Sainio,~S.; Laurila,~T.; Caro,~M.~A. Understanding
  {{X-ray Spectroscopy}} of {{Carbonaceous Materials}} by {{Combining
  Experiments}}, {{Density Functional Theory}}, and {{Machine Learning}}.
  {{Part II}}: {{Quantitative Fitting}} of {{Spectra}}. \emph{Chem. Mater.}
  \textbf{2019}, \emph{31}, 9256--9267\relax
\mciteBstWouldAddEndPuncttrue
\mciteSetBstMidEndSepPunct{\mcitedefaultmidpunct}
{\mcitedefaultendpunct}{\mcitedefaultseppunct}\relax
\EndOfBibitem
\bibitem[Zheng and Cheng(2019)Zheng, and
  Cheng]{zhengPerformanceDeltaCoupledClusterMethods2019}
Zheng,~X.; Cheng,~L. Performance of {{Delta-Coupled-Cluster Methods}} for
  {{Calculations}} of {{Core-Ionization Energies}} of {{First-Row Elements}}.
  \emph{J. Chem. Theory Comput.} \textbf{2019}, \emph{15}, 4945--4955\relax
\mciteBstWouldAddEndPuncttrue
\mciteSetBstMidEndSepPunct{\mcitedefaultmidpunct}
{\mcitedefaultendpunct}{\mcitedefaultseppunct}\relax
\EndOfBibitem
\bibitem[Sen \latin{et~al.}(2018)Sen, Shee, and
  Mukherjee]{senInclusionOrbitalRelaxation2018}
Sen,~S.; Shee,~A.; Mukherjee,~D. Inclusion of Orbital Relaxation and
  Correlation through the Unitary Group Adapted Open Shell Coupled Cluster
  Theory Using Non-Relativistic and Scalar Relativistic {{Hamiltonians}} to
  Study the Core Ionization Potential of Molecules Containing Light to
  Medium-Heavy Elements. \emph{J. Chem. Phys.} \textbf{2018}, \emph{148},
  054107\relax
\mciteBstWouldAddEndPuncttrue
\mciteSetBstMidEndSepPunct{\mcitedefaultmidpunct}
{\mcitedefaultendpunct}{\mcitedefaultseppunct}\relax
\EndOfBibitem
\bibitem[Hohenberg and Kohn(1964)Hohenberg, and
  Kohn]{hohenbergInhomogeneousElectronGas1964}
Hohenberg,~P.; Kohn,~W. Inhomogeneous {{Electron Gas}}. \emph{Phys. Rev.}
  \textbf{1964}, \emph{136}, B864--B871\relax
\mciteBstWouldAddEndPuncttrue
\mciteSetBstMidEndSepPunct{\mcitedefaultmidpunct}
{\mcitedefaultendpunct}{\mcitedefaultseppunct}\relax
\EndOfBibitem
\bibitem[Kohn and Sham(1965)Kohn, and
  Sham]{kohnSelfConsistentEquationsIncluding1965}
Kohn,~W.; Sham,~L.~J. Self-{{Consistent Equations Including Exchange}} and
  {{Correlation Effects}}. \emph{Phys. Rev.} \textbf{1965}, \emph{140},
  A1133--A1138\relax
\mciteBstWouldAddEndPuncttrue
\mciteSetBstMidEndSepPunct{\mcitedefaultmidpunct}
{\mcitedefaultendpunct}{\mcitedefaultseppunct}\relax
\EndOfBibitem
\bibitem[Parr and Yang(1989)Parr, and
  Yang]{parrDensityFunctionalTheoryAtoms1989}
Parr,~R.~G.; Yang,~W. \emph{Density-{{Functional Theory}} of {{Atoms}} and
  {{Molecules}}}; {Oxford University Press}, 1989\relax
\mciteBstWouldAddEndPuncttrue
\mciteSetBstMidEndSepPunct{\mcitedefaultmidpunct}
{\mcitedefaultendpunct}{\mcitedefaultseppunct}\relax
\EndOfBibitem
\bibitem[Bagus(1965)]{bagusSelfConsistentFieldWaveFunctions1965}
Bagus,~P.~S. Self-{{Consistent-Field Wave Functions}} for {{Hole States}} of
  {{Some Ne-Like}} and {{Ar-Like Ions}}. \emph{Phys. Rev.} \textbf{1965},
  \emph{139}, A619--A634\relax
\mciteBstWouldAddEndPuncttrue
\mciteSetBstMidEndSepPunct{\mcitedefaultmidpunct}
{\mcitedefaultendpunct}{\mcitedefaultseppunct}\relax
\EndOfBibitem
\bibitem[Pueyo~Bellafont \latin{et~al.}(2015)Pueyo~Bellafont, Bagus, and
  Illas]{pueyobellafontPredictionCoreLevel2015}
Pueyo~Bellafont,~N.; Bagus,~P.~S.; Illas,~F. Prediction of Core Level Binding
  Energies in Density Functional Theory: {{Rigorous}} Definition of Initial and
  Final State Contributions and Implications on the Physical Meaning of
  {{Kohn-Sham}} Energies. \emph{J. Chem. Phys.} \textbf{2015}, \emph{142},
  214102\relax
\mciteBstWouldAddEndPuncttrue
\mciteSetBstMidEndSepPunct{\mcitedefaultmidpunct}
{\mcitedefaultendpunct}{\mcitedefaultseppunct}\relax
\EndOfBibitem
\bibitem[Pueyo~Bellafont \latin{et~al.}(2016)Pueyo~Bellafont, {\'A}lvarez~Saiz,
  Vi{\~n}es, and Illas]{pueyobellafontPerformanceMinnesotaFunctionals2016}
Pueyo~Bellafont,~N.; {\'A}lvarez~Saiz,~G.; Vi{\~n}es,~F.; Illas,~F. Performance
  of {{Minnesota}} Functionals on Predicting Core-Level Binding Energies of
  Molecules Containing Main-Group Elements. \emph{Theor Chem Acc}
  \textbf{2016}, \emph{135}, 35\relax
\mciteBstWouldAddEndPuncttrue
\mciteSetBstMidEndSepPunct{\mcitedefaultmidpunct}
{\mcitedefaultendpunct}{\mcitedefaultseppunct}\relax
\EndOfBibitem
\bibitem[Vi{\~n}es \latin{et~al.}(2018)Vi{\~n}es, Sousa, and
  Illas]{vinesPredictionCoreLevel2018}
Vi{\~n}es,~F.; Sousa,~C.; Illas,~F. On the Prediction of Core Level Binding
  Energies in Molecules, Surfaces and Solids. \emph{Phys. Chem. Chem. Phys.}
  \textbf{2018}, \emph{20}, 8403--8410\relax
\mciteBstWouldAddEndPuncttrue
\mciteSetBstMidEndSepPunct{\mcitedefaultmidpunct}
{\mcitedefaultendpunct}{\mcitedefaultseppunct}\relax
\EndOfBibitem
\bibitem[Kahk and Lischner(2019)Kahk, and
  Lischner]{kahkAccurateAbsoluteCoreelectron2019}
Kahk,~J.~M.; Lischner,~J. Accurate Absolute Core-Electron Binding Energies of
  Molecules, Solids, and Surfaces from First-Principles Calculations.
  \emph{Phys. Rev. Mater.} \textbf{2019}, \emph{3}, 100801\relax
\mciteBstWouldAddEndPuncttrue
\mciteSetBstMidEndSepPunct{\mcitedefaultmidpunct}
{\mcitedefaultendpunct}{\mcitedefaultseppunct}\relax
\EndOfBibitem
\bibitem[Klein \latin{et~al.}(2021)Klein, Hall, and
  Maurer]{kleinNutsBoltsCorehole2021}
Klein,~B.~P.; Hall,~S.~J.; Maurer,~R.~J. The Nuts and Bolts of Core-Hole
  Constrained Ab Initio Simulation for {{K-shell}} x-Ray Photoemission and
  Absorption Spectra. \emph{J. Phys.: Condens. Matter} \textbf{2021},
  \emph{33}, 154005\relax
\mciteBstWouldAddEndPuncttrue
\mciteSetBstMidEndSepPunct{\mcitedefaultmidpunct}
{\mcitedefaultendpunct}{\mcitedefaultseppunct}\relax
\EndOfBibitem
\bibitem[Kahk \latin{et~al.}(2021)Kahk, Michelitsch, Maurer, Reuter, and
  Lischner]{kahkCoreElectronBinding2021}
Kahk,~J.~M.; Michelitsch,~G.~S.; Maurer,~R.~J.; Reuter,~K.; Lischner,~J. Core
  {{Electron Binding Energies}} in {{Solids}} from {{Periodic All-Electron
  $\Delta$-Self-Consistent-Field Calculations}}. \emph{J. Phys. Chem. Lett.}
  \textbf{2021}, \emph{12}, 9353--9359\relax
\mciteBstWouldAddEndPuncttrue
\mciteSetBstMidEndSepPunct{\mcitedefaultmidpunct}
{\mcitedefaultendpunct}{\mcitedefaultseppunct}\relax
\EndOfBibitem
\bibitem[Pinheiro \latin{et~al.}(2015)Pinheiro, Caldas, Rinke, Blum, and
  Scheffler]{pinheiroLengthDependenceIonization2015}
Pinheiro,~M.; Caldas,~M.~J.; Rinke,~P.; Blum,~V.; Scheffler,~M. Length
  Dependence of Ionization Potentials of Transacetylenes: {{Internally}}
  Consistent {{DFT}}/${{GW}}$ Approach. \emph{Phys. Rev. B} \textbf{2015},
  \emph{92}, 195134\relax
\mciteBstWouldAddEndPuncttrue
\mciteSetBstMidEndSepPunct{\mcitedefaultmidpunct}
{\mcitedefaultendpunct}{\mcitedefaultseppunct}\relax
\EndOfBibitem
\bibitem[Golze \latin{et~al.}(2018)Golze, Wilhelm, {van Setten}, and
  Rinke]{golzeCoreLevelBindingEnergies2018}
Golze,~D.; Wilhelm,~J.; {van Setten},~M.~J.; Rinke,~P. Core-{{Level Binding
  Energies}} from {{GW}}: {{An Efficient Full-Frequency Approach}} within a
  {{Localized Basis}}. \emph{J. Chem. Theory Comput.} \textbf{2018}, \emph{14},
  4856--4869\relax
\mciteBstWouldAddEndPuncttrue
\mciteSetBstMidEndSepPunct{\mcitedefaultmidpunct}
{\mcitedefaultendpunct}{\mcitedefaultseppunct}\relax
\EndOfBibitem
\bibitem[Michelitsch and Reuter(2019)Michelitsch, and
  Reuter]{michelitschEfficientSimulationNearedge2019}
Michelitsch,~G.~S.; Reuter,~K. Efficient Simulation of Near-Edge x-Ray
  Absorption Fine Structure ({{NEXAFS}}) in Density-Functional Theory:
  {{Comparison}} of Core-Level Constraining Approaches. \emph{J. Chem. Phys.}
  \textbf{2019}, \emph{150}, 074104\relax
\mciteBstWouldAddEndPuncttrue
\mciteSetBstMidEndSepPunct{\mcitedefaultmidpunct}
{\mcitedefaultendpunct}{\mcitedefaultseppunct}\relax
\EndOfBibitem
\bibitem[Golze \latin{et~al.}(2021)Golze, Hirvensalo, {Hern{\'a}ndez-Le{\'o}n},
  Aarva, Etula, Susi, Rinke, Laurila, and
  Caro]{golzeAccurateComputationalPrediction2021}
Golze,~D.; Hirvensalo,~M.; {Hern{\'a}ndez-Le{\'o}n},~P.; Aarva,~A.; Etula,~J.;
  Susi,~T.; Rinke,~P.; Laurila,~T.; Caro,~M.~A. Accurate Computational
  Prediction of Core-Electron Binding Energies in Carbon-Based Materials: {{A}}
  Machine-Learning Model Combining {{DFT}} and $GW$. \emph{arXiv:2112.06551}
  \textbf{2021}, \relax
\mciteBstWouldAddEndPunctfalse
\mciteSetBstMidEndSepPunct{\mcitedefaultmidpunct}
{}{\mcitedefaultseppunct}\relax
\EndOfBibitem
\bibitem[Hall \latin{et~al.}(2021)Hall, Klein, and
  Maurer]{hallSelfinteractionErrorInduces2021}
Hall,~S.~J.; Klein,~B.~P.; Maurer,~R.~J. Self-Interaction Error Induces
  Spurious Charge Transfer Artefacts in Core-Level Simulations of x-Ray
  Photoemission and Absorption Spectroscopy of Metal-Organic Interfaces.
  \emph{arXiv:2112.00876} \textbf{2021}, \relax
\mciteBstWouldAddEndPunctfalse
\mciteSetBstMidEndSepPunct{\mcitedefaultmidpunct}
{}{\mcitedefaultseppunct}\relax
\EndOfBibitem
\bibitem[Liu \latin{et~al.}(2019)Liu, Matthews, Coriani, and
  Cheng]{liuBenchmarkCalculationsKEdge2019}
Liu,~J.; Matthews,~D.; Coriani,~S.; Cheng,~L. Benchmark {{Calculations}} of
  {{K-Edge Ionization Energies}} for {{First-Row Elements Using
  Scalar-Relativistic Core}}\textendash{{Valence-Separated Equation-of-Motion
  Coupled-Cluster Methods}}. \emph{J. Chem. Theory Comput.} \textbf{2019},
  \emph{15}, 1642--1651\relax
\mciteBstWouldAddEndPuncttrue
\mciteSetBstMidEndSepPunct{\mcitedefaultmidpunct}
{\mcitedefaultendpunct}{\mcitedefaultseppunct}\relax
\EndOfBibitem
\bibitem[Vidal \latin{et~al.}(2020)Vidal, Pokhilko, Krylov, and
  Coriani]{vidalEquationofMotionCoupledClusterTheory2020}
Vidal,~M.~L.; Pokhilko,~P.; Krylov,~A.~I.; Coriani,~S. Equation-of-{{Motion
  Coupled-Cluster Theory}} to {{Model L-Edge X-ray Absorption}} and
  {{Photoelectron Spectra}}. \emph{J. Phys. Chem. Lett.} \textbf{2020},
  \emph{11}, 8314--8321\relax
\mciteBstWouldAddEndPuncttrue
\mciteSetBstMidEndSepPunct{\mcitedefaultmidpunct}
{\mcitedefaultendpunct}{\mcitedefaultseppunct}\relax
\EndOfBibitem
\bibitem[Ambroise \latin{et~al.}(2021)Ambroise, Dreuw, and
  Jensen]{ambroiseProbingBasisSet2021}
Ambroise,~M.~A.; Dreuw,~A.; Jensen,~F. Probing {{Basis Set Requirements}} for
  {{Calculating Core Ionization}} and {{Core Excitation Spectra Using
  Correlated Wave Function Methods}}. \emph{J. Chem. Theory Comput.}
  \textbf{2021}, \emph{17}, 2832--2842\relax
\mciteBstWouldAddEndPuncttrue
\mciteSetBstMidEndSepPunct{\mcitedefaultmidpunct}
{\mcitedefaultendpunct}{\mcitedefaultseppunct}\relax
\EndOfBibitem
\bibitem[Vila \latin{et~al.}(2021)Vila, Kas, Rehr, Kowalski, and
  Peng]{vilaEquationofMotionCoupledClusterCumulant2021}
Vila,~F.~D.; Kas,~J.~J.; Rehr,~J.~J.; Kowalski,~K.; Peng,~B.
  Equation-of-{{Motion Coupled-Cluster Cumulant Green}}'s {{Function}} for
  {{Excited States}} and {{X-Ray Spectra}}. \emph{Front. Chem.} \textbf{2021},
  \emph{9}\relax
\mciteBstWouldAddEndPuncttrue
\mciteSetBstMidEndSepPunct{\mcitedefaultmidpunct}
{\mcitedefaultendpunct}{\mcitedefaultseppunct}\relax
\EndOfBibitem
\bibitem[Martin \latin{et~al.}(2016)Martin, Reining, and
  Ceperley]{martinInteractingElectrons2016}
Martin,~R.~M.; Reining,~L.; Ceperley,~D.~M. \emph{Interacting {{Electrons}}};
  {Cambridge University Press}, 2016\relax
\mciteBstWouldAddEndPuncttrue
\mciteSetBstMidEndSepPunct{\mcitedefaultmidpunct}
{\mcitedefaultendpunct}{\mcitedefaultseppunct}\relax
\EndOfBibitem
\bibitem[Golze \latin{et~al.}(2019)Golze, Dvorak, and
  Rinke]{golzeGWCompendiumPractical2019}
Golze,~D.; Dvorak,~M.; Rinke,~P. The {{GW Compendium}}: {{A Practical Guide}}
  to {{Theoretical Photoemission Spectroscopy}}. \emph{Front. Chem.}
  \textbf{2019}, \emph{7}\relax
\mciteBstWouldAddEndPuncttrue
\mciteSetBstMidEndSepPunct{\mcitedefaultmidpunct}
{\mcitedefaultendpunct}{\mcitedefaultseppunct}\relax
\EndOfBibitem
\bibitem[Reining(2018)]{reiningGWApproximationContent2018}
Reining,~L. The {{GW}} Approximation: Content, Successes and Limitations.
  \emph{WIREs Comput. Mol. Sci} \textbf{2018}, \emph{8}, e1344\relax
\mciteBstWouldAddEndPuncttrue
\mciteSetBstMidEndSepPunct{\mcitedefaultmidpunct}
{\mcitedefaultendpunct}{\mcitedefaultseppunct}\relax
\EndOfBibitem
\bibitem[Hedin(1965)]{hedinNewMethodCalculating1965}
Hedin,~L. New {{Method}} for {{Calculating}} the {{One-Particle Green}}'s
  {{Function}} with {{Application}} to the {{Electron-Gas Problem}}.
  \emph{Phys. Rev.} \textbf{1965}, \emph{139}, A796--A823\relax
\mciteBstWouldAddEndPuncttrue
\mciteSetBstMidEndSepPunct{\mcitedefaultmidpunct}
{\mcitedefaultendpunct}{\mcitedefaultseppunct}\relax
\EndOfBibitem
\bibitem[Rasmussen \latin{et~al.}(2021)Rasmussen, Deilmann, and
  Thygesen]{rasmussenFullyAutomatedGW2021}
Rasmussen,~A.; Deilmann,~T.; Thygesen,~K.~S. Towards Fully Automated {{GW}}
  Band Structure Calculations: {{What}} We Can Learn from 60.000 Self-Energy
  Evaluations. \emph{npj Comput Mater} \textbf{2021}, \emph{7}, 1--9\relax
\mciteBstWouldAddEndPuncttrue
\mciteSetBstMidEndSepPunct{\mcitedefaultmidpunct}
{\mcitedefaultendpunct}{\mcitedefaultseppunct}\relax
\EndOfBibitem
\bibitem[{van Setten} \latin{et~al.}(2015){van Setten}, Caruso, Sharifzadeh,
  Ren, Scheffler, Liu, Lischner, Lin, Deslippe, Louie, Yang, Weigend, Neaton,
  Evers, and Rinke]{vansettenGW100BenchmarkingG0W02015}
{van Setten},~M.~J.; Caruso,~F.; Sharifzadeh,~S.; Ren,~X.; Scheffler,~M.;
  Liu,~F.; Lischner,~J.; Lin,~L.; Deslippe,~J.~R.; Louie,~S.~G.; Yang,~C.;
  Weigend,~F.; Neaton,~J.~B.; Evers,~F.; Rinke,~P. {{GW100}}: {{Benchmarking
  G0W0}} for {{Molecular Systems}}. \emph{J. Chem. Theory Comput.}
  \textbf{2015}, \emph{11}, 5665--5687\relax
\mciteBstWouldAddEndPuncttrue
\mciteSetBstMidEndSepPunct{\mcitedefaultmidpunct}
{\mcitedefaultendpunct}{\mcitedefaultseppunct}\relax
\EndOfBibitem
\bibitem[Stuke \latin{et~al.}(2020)Stuke, Kunkel, Golze, Todorovi{\'c},
  Margraf, Reuter, Rinke, and Oberhofer]{stukeAtomicStructuresOrbital2020}
Stuke,~A.; Kunkel,~C.; Golze,~D.; Todorovi{\'c},~M.; Margraf,~J.~T.;
  Reuter,~K.; Rinke,~P.; Oberhofer,~H. Atomic Structures and Orbital Energies
  of 61,489 Crystal-Forming Organic Molecules. \emph{Sci Data} \textbf{2020},
  \emph{7}, 58\relax
\mciteBstWouldAddEndPuncttrue
\mciteSetBstMidEndSepPunct{\mcitedefaultmidpunct}
{\mcitedefaultendpunct}{\mcitedefaultseppunct}\relax
\EndOfBibitem
\bibitem[Bruneval \latin{et~al.}(2021)Bruneval, Dattani, and {van
  Setten}]{brunevalGWMiracleManyBody2021}
Bruneval,~F.; Dattani,~N.; {van Setten},~M.~J. The {{GW Miracle}} in
  {{Many-Body Perturbation Theory}} for the {{Ionization Potential}} of
  {{Molecules}}. \emph{Front. Chem.} \textbf{2021}, \emph{9}\relax
\mciteBstWouldAddEndPuncttrue
\mciteSetBstMidEndSepPunct{\mcitedefaultmidpunct}
{\mcitedefaultendpunct}{\mcitedefaultseppunct}\relax
\EndOfBibitem
\bibitem[Blase \latin{et~al.}(2011)Blase, Attaccalite, and
  Olevano]{blaseFirstprinciplesMathitGWCalculations2011}
Blase,~X.; Attaccalite,~C.; Olevano,~V. First-Principles $GW$ Calculations for
  Fullerenes, Porphyrins, Phtalocyanine, and Other Molecules of Interest for
  Organic Photovoltaic Applications. \emph{Phys. Rev. B} \textbf{2011},
  \emph{83}, 115103\relax
\mciteBstWouldAddEndPuncttrue
\mciteSetBstMidEndSepPunct{\mcitedefaultmidpunct}
{\mcitedefaultendpunct}{\mcitedefaultseppunct}\relax
\EndOfBibitem
\bibitem[Ren \latin{et~al.}(2012)Ren, Rinke, Blum, Wieferink, Tkatchenko,
  Sanfilippo, Reuter, and
  Scheffler]{renResolutionofidentityApproachHartree2012}
Ren,~X.; Rinke,~P.; Blum,~V.; Wieferink,~J.; Tkatchenko,~A.; Sanfilippo,~A.;
  Reuter,~K.; Scheffler,~M. Resolution-of-Identity Approach to
  {{Hartree}}\textendash{{Fock}}, Hybrid Density Functionals, {{RPA}}, {{MP2
  andGWwith}} Numeric Atom-Centered Orbital Basis Functions. \emph{New J.
  Phys.} \textbf{2012}, \emph{14}, 053020\relax
\mciteBstWouldAddEndPuncttrue
\mciteSetBstMidEndSepPunct{\mcitedefaultmidpunct}
{\mcitedefaultendpunct}{\mcitedefaultseppunct}\relax
\EndOfBibitem
\bibitem[{van Setten} \latin{et~al.}(2013){van Setten}, Weigend, and
  Evers]{vansettenGWMethodQuantumChemistry2013}
{van Setten},~M.~J.; Weigend,~F.; Evers,~F. The {{GW-Method}} for {{Quantum
  Chemistry Applications}}: {{Theory}} and {{Implementation}}. \emph{J. Chem.
  Theory Comput.} \textbf{2013}, \emph{9}, 232--246\relax
\mciteBstWouldAddEndPuncttrue
\mciteSetBstMidEndSepPunct{\mcitedefaultmidpunct}
{\mcitedefaultendpunct}{\mcitedefaultseppunct}\relax
\EndOfBibitem
\bibitem[Bruneval \latin{et~al.}(2016)Bruneval, Rangel, Hamed, Shao, Yang, and
  Neaton]{brunevalMolgwManybodyPerturbation2016}
Bruneval,~F.; Rangel,~T.; Hamed,~S.~M.; Shao,~M.; Yang,~C.; Neaton,~J.~B. Molgw
  1: {{Many-body}} Perturbation Theory Software for Atoms, Molecules, and
  Clusters. \emph{Comput. Phys. Commun} \textbf{2016}, \emph{208},
  149--161\relax
\mciteBstWouldAddEndPuncttrue
\mciteSetBstMidEndSepPunct{\mcitedefaultmidpunct}
{\mcitedefaultendpunct}{\mcitedefaultseppunct}\relax
\EndOfBibitem
\bibitem[Wilhelm \latin{et~al.}(2016)Wilhelm, Del~Ben, and
  Hutter]{wilhelmGWGaussianPlane2016}
Wilhelm,~J.; Del~Ben,~M.; Hutter,~J. {{GW}} in the {{Gaussian}} and {{Plane
  Waves Scheme}} with {{Application}} to {{Linear Acenes}}. \emph{J. Chem.
  Theory Comput.} \textbf{2016}, \emph{12}, 3623--3635\relax
\mciteBstWouldAddEndPuncttrue
\mciteSetBstMidEndSepPunct{\mcitedefaultmidpunct}
{\mcitedefaultendpunct}{\mcitedefaultseppunct}\relax
\EndOfBibitem
\bibitem[F{\"o}rster \latin{et~al.}(2020)F{\"o}rster, Franchini, {van Lenthe},
  and Visscher]{forsterQuadraticPairAtomic2020}
F{\"o}rster,~A.; Franchini,~M.; {van Lenthe},~E.; Visscher,~L. A {{Quadratic
  Pair Atomic Resolution}} of the {{Identity Based SOS-AO-MP2 Algorithm Using
  Slater Type Orbitals}}. \emph{J. Chem. Theory Comput.} \textbf{2020},
  \emph{16}, 875--891\relax
\mciteBstWouldAddEndPuncttrue
\mciteSetBstMidEndSepPunct{\mcitedefaultmidpunct}
{\mcitedefaultendpunct}{\mcitedefaultseppunct}\relax
\EndOfBibitem
\bibitem[Zhu and Chan(2021)Zhu, and Chan]{zhuAllElectronGaussianBasedG0W02021}
Zhu,~T.; Chan,~G. K.-L. All-{{Electron Gaussian-Based G0W0}} for {{Valence}}
  and {{Core Excitation Energies}} of {{Periodic Systems}}. \emph{J. Chem.
  Theory Comput.} \textbf{2021}, \emph{17}, 727--741\relax
\mciteBstWouldAddEndPuncttrue
\mciteSetBstMidEndSepPunct{\mcitedefaultmidpunct}
{\mcitedefaultendpunct}{\mcitedefaultseppunct}\relax
\EndOfBibitem
\bibitem[{Mejia-Rodriguez} \latin{et~al.}(2021){Mejia-Rodriguez}, Kunitsa,
  Apr{\`a}, and Govind]{mejia-rodriguezScalableMolecularGW2021}
{Mejia-Rodriguez},~D.; Kunitsa,~A.; Apr{\`a},~E.; Govind,~N. Scalable
  {{Molecular GW Calculations}}: {{Valence}} and {{Core Spectra}}. \emph{J.
  Chem. Theory Comput.} \textbf{2021}, \emph{17}, 7504--7517\relax
\mciteBstWouldAddEndPuncttrue
\mciteSetBstMidEndSepPunct{\mcitedefaultmidpunct}
{\mcitedefaultendpunct}{\mcitedefaultseppunct}\relax
\EndOfBibitem
\bibitem[Aoki and Ohno(2018)Aoki, and
  Ohno]{aokiAccurateQuasiparticleCalculation2018}
Aoki,~T.; Ohno,~K. Accurate Quasiparticle Calculation of X-Ray Photoelectron
  Spectra of Solids. \emph{J. Phys.: Condens. Matter} \textbf{2018}, \emph{30},
  21LT01\relax
\mciteBstWouldAddEndPuncttrue
\mciteSetBstMidEndSepPunct{\mcitedefaultmidpunct}
{\mcitedefaultendpunct}{\mcitedefaultseppunct}\relax
\EndOfBibitem
\bibitem[{van Setten} \latin{et~al.}(2018){van Setten}, Costa, Vi{\~n}es, and
  Illas]{vansettenAssessingGWApproaches2018}
{van Setten},~M.~J.; Costa,~R.; Vi{\~n}es,~F.; Illas,~F. Assessing {{GW
  Approaches}} for {{Predicting Core Level Binding Energies}}. \emph{J. Chem.
  Theory Comput.} \textbf{2018}, \emph{14}, 877--883\relax
\mciteBstWouldAddEndPuncttrue
\mciteSetBstMidEndSepPunct{\mcitedefaultmidpunct}
{\mcitedefaultendpunct}{\mcitedefaultseppunct}\relax
\EndOfBibitem
\bibitem[Voora \latin{et~al.}(2019)Voora, Galhenage, Hemminger, and
  Furche]{vooraEffectiveOneparticleEnergies2019}
Voora,~V.~K.; Galhenage,~R.; Hemminger,~J.~C.; Furche,~F. Effective
  One-Particle Energies from Generalized {{Kohn}}\textendash{{Sham}} Random
  Phase Approximation: {{A}} Direct Approach for Computing and Analyzing Core
  Ionization Energies. \emph{J. Chem. Phys.} \textbf{2019}, \emph{151},
  134106\relax
\mciteBstWouldAddEndPuncttrue
\mciteSetBstMidEndSepPunct{\mcitedefaultmidpunct}
{\mcitedefaultendpunct}{\mcitedefaultseppunct}\relax
\EndOfBibitem
\bibitem[Golze \latin{et~al.}(2020)Golze, Keller, and
  Rinke]{golzeAccurateAbsoluteRelative2020}
Golze,~D.; Keller,~L.; Rinke,~P. Accurate {{Absolute}} and {{Relative
  Core-Level Binding Energies}} from {{GW}}. \emph{J. Phys. Chem. Lett.}
  \textbf{2020}, \emph{11}, 1840--1847\relax
\mciteBstWouldAddEndPuncttrue
\mciteSetBstMidEndSepPunct{\mcitedefaultmidpunct}
{\mcitedefaultendpunct}{\mcitedefaultseppunct}\relax
\EndOfBibitem
\bibitem[Keller \latin{et~al.}(2020)Keller, Blum, Rinke, and
  Golze]{kellerRelativisticCorrectionScheme2020}
Keller,~L.; Blum,~V.; Rinke,~P.; Golze,~D. Relativistic Correction Scheme for
  Core-Level Binding Energies from {{GW}}. \emph{J. Chem. Phys.} \textbf{2020},
  \emph{153}, 114110\relax
\mciteBstWouldAddEndPuncttrue
\mciteSetBstMidEndSepPunct{\mcitedefaultmidpunct}
{\mcitedefaultendpunct}{\mcitedefaultseppunct}\relax
\EndOfBibitem
\bibitem[Duchemin and Blase(2020)Duchemin, and
  Blase]{ducheminRobustAnalyticContinuationApproach2020}
Duchemin,~I.; Blase,~X. Robust {{Analytic-Continuation Approach}} to
  {{Many-Body GW Calculations}}. \emph{J. Chem. Theory Comput.} \textbf{2020},
  \emph{16}, 1742--1756\relax
\mciteBstWouldAddEndPuncttrue
\mciteSetBstMidEndSepPunct{\mcitedefaultmidpunct}
{\mcitedefaultendpunct}{\mcitedefaultseppunct}\relax
\EndOfBibitem
\bibitem[{Mejia-Rodriguez} \latin{et~al.}(2022){Mejia-Rodriguez}, Kunitsa,
  Apr{\`a}, and Govind]{mejia-rodriguezBasisSetSelection2022}
{Mejia-Rodriguez},~D.; Kunitsa,~A.; Apr{\`a},~E.; Govind,~N. On the Basis Set
  Selection for Molecular Core-Level ${{GW}}$ Calculations.
  \emph{arXiv:2203.10169} \textbf{2022}, \relax
\mciteBstWouldAddEndPunctfalse
\mciteSetBstMidEndSepPunct{\mcitedefaultmidpunct}
{}{\mcitedefaultseppunct}\relax
\EndOfBibitem
\bibitem[Yao \latin{et~al.}(2022)Yao, Golze, Rinke, Blum, and
  Kanai]{yaoAllElectronBSEGW2022}
Yao,~Y.; Golze,~D.; Rinke,~P.; Blum,~V.; Kanai,~Y. All-{{Electron BSE}}@{{GW
  Method}} for {{K-Edge Core Electron Excitation Energies}}. \emph{J. Chem.
  Theory Comput.} \textbf{2022}, \relax
\mciteBstWouldAddEndPunctfalse
\mciteSetBstMidEndSepPunct{\mcitedefaultmidpunct}
{}{\mcitedefaultseppunct}\relax
\EndOfBibitem
\bibitem[Wilhelm and Hutter(2017)Wilhelm, and
  Hutter]{wilhelmPeriodicGWCalculations2017}
Wilhelm,~J.; Hutter,~J. Periodic ${{GW}}$ Calculations in the {{Gaussian}} and
  Plane-Waves Scheme. \emph{Phys. Rev. B} \textbf{2017}, \emph{95},
  235123\relax
\mciteBstWouldAddEndPuncttrue
\mciteSetBstMidEndSepPunct{\mcitedefaultmidpunct}
{\mcitedefaultendpunct}{\mcitedefaultseppunct}\relax
\EndOfBibitem
\bibitem[Ren \latin{et~al.}(2021)Ren, Merz, Jiang, Yao, Rampp, Lederer, Blum,
  and Scheffler]{renAllelectronPeriodic0W2021}
Ren,~X.; Merz,~F.; Jiang,~H.; Yao,~Y.; Rampp,~M.; Lederer,~H.; Blum,~V.;
  Scheffler,~M. All-Electron Periodic $G_0W_0$ Implementation with Numerical
  Atomic Orbital Basis Functions: {{Algorithm}} and Benchmarks. \emph{Phys.
  Rev. Mater.} \textbf{2021}, \emph{5}, 013807\relax
\mciteBstWouldAddEndPuncttrue
\mciteSetBstMidEndSepPunct{\mcitedefaultmidpunct}
{\mcitedefaultendpunct}{\mcitedefaultseppunct}\relax
\EndOfBibitem
\bibitem[Wilhelm \latin{et~al.}(2018)Wilhelm, Golze, Talirz, Hutter, and
  Pignedoli]{wilhelmGWCalculationsThousands2018}
Wilhelm,~J.; Golze,~D.; Talirz,~L.; Hutter,~J.; Pignedoli,~C.~A. Toward {{GW
  Calculations}} on {{Thousands}} of {{Atoms}}. \emph{J. Phys. Chem. Lett.}
  \textbf{2018}, \emph{9}, 306--312\relax
\mciteBstWouldAddEndPuncttrue
\mciteSetBstMidEndSepPunct{\mcitedefaultmidpunct}
{\mcitedefaultendpunct}{\mcitedefaultseppunct}\relax
\EndOfBibitem
\bibitem[Wilhelm \latin{et~al.}(2021)Wilhelm, Seewald, and
  Golze]{wilhelmLowScalingGWBenchmark2021}
Wilhelm,~J.; Seewald,~P.; Golze,~D. Low-{{Scaling GW}} with {{Benchmark
  Accuracy}} and {{Application}} to {{Phosphorene Nanosheets}}. \emph{J. Chem.
  Theory Comput.} \textbf{2021}, \emph{17}, 1662--1677\relax
\mciteBstWouldAddEndPuncttrue
\mciteSetBstMidEndSepPunct{\mcitedefaultmidpunct}
{\mcitedefaultendpunct}{\mcitedefaultseppunct}\relax
\EndOfBibitem
\bibitem[Duchemin and Blase(2021)Duchemin, and
  Blase]{ducheminCubicScalingAllElectronGW2021}
Duchemin,~I.; Blase,~X. Cubic-{{Scaling All-Electron GW Calculations}} with a
  {{Separable Density-Fitting Space}}\textendash{{Time Approach}}. \emph{J.
  Chem. Theory Comput.} \textbf{2021}, \emph{17}, 2383--2393\relax
\mciteBstWouldAddEndPuncttrue
\mciteSetBstMidEndSepPunct{\mcitedefaultmidpunct}
{\mcitedefaultendpunct}{\mcitedefaultseppunct}\relax
\EndOfBibitem
\bibitem[F{\"o}rster and Visscher(2021)F{\"o}rster, and
  Visscher]{forsterLowOrderScalingQuasiparticle2021}
F{\"o}rster,~A.; Visscher,~L. Low-{{Order Scaling Quasiparticle Self-Consistent
  GW}} for {{Molecules}}. \emph{Front. Chem.} \textbf{2021}, \emph{9}\relax
\mciteBstWouldAddEndPuncttrue
\mciteSetBstMidEndSepPunct{\mcitedefaultmidpunct}
{\mcitedefaultendpunct}{\mcitedefaultseppunct}\relax
\EndOfBibitem
\bibitem[Sch{\"o}ne and Eguiluz(1998)Sch{\"o}ne, and
  Eguiluz]{schoneSelfConsistentCalculationsQuasiparticle1998}
Sch{\"o}ne,~W.-D.; Eguiluz,~A.~G. Self-{{Consistent Calculations}} of
  {{Quasiparticle States}} in {{Metals}} and {{Semiconductors}}. \emph{Phys.
  Rev. Lett.} \textbf{1998}, \emph{81}, 1662--1665\relax
\mciteBstWouldAddEndPuncttrue
\mciteSetBstMidEndSepPunct{\mcitedefaultmidpunct}
{\mcitedefaultendpunct}{\mcitedefaultseppunct}\relax
\EndOfBibitem
\bibitem[Caruso \latin{et~al.}(2012)Caruso, Rinke, Ren, Scheffler, and
  Rubio]{carusoUnifiedDescriptionGround2012}
Caruso,~F.; Rinke,~P.; Ren,~X.; Scheffler,~M.; Rubio,~A. Unified Description of
  Ground and Excited States of Finite Systems: {{The}} Self-Consistent ${{GW}}$
  Approach. \emph{Phys. Rev. B} \textbf{2012}, \emph{86}, 081102\relax
\mciteBstWouldAddEndPuncttrue
\mciteSetBstMidEndSepPunct{\mcitedefaultmidpunct}
{\mcitedefaultendpunct}{\mcitedefaultseppunct}\relax
\EndOfBibitem
\bibitem[Caruso \latin{et~al.}(2013)Caruso, Rinke, Ren, Rubio, and
  Scheffler]{carusoSelfconsistentGWAllelectron2013}
Caruso,~F.; Rinke,~P.; Ren,~X.; Rubio,~A.; Scheffler,~M. Self-Consistent
  {{GW}}: {{All-electron}} Implementation with Localized Basis Functions.
  \emph{Phys. Rev. B} \textbf{2013}, \emph{88}, 075105\relax
\mciteBstWouldAddEndPuncttrue
\mciteSetBstMidEndSepPunct{\mcitedefaultmidpunct}
{\mcitedefaultendpunct}{\mcitedefaultseppunct}\relax
\EndOfBibitem
\bibitem[{van~Schilfgaarde} \latin{et~al.}(2006){van~Schilfgaarde}, Kotani, and
  Faleev]{vanschilfgaardeQuasiparticleSelfConsistentGW2006}
{van~Schilfgaarde},~M.; Kotani,~T.; Faleev,~S. Quasiparticle
  {{Self-Consistent}} ${{GW}}$ {{Theory}}. \emph{Phys. Rev. Lett.}
  \textbf{2006}, \emph{96}, 226402\relax
\mciteBstWouldAddEndPuncttrue
\mciteSetBstMidEndSepPunct{\mcitedefaultmidpunct}
{\mcitedefaultendpunct}{\mcitedefaultseppunct}\relax
\EndOfBibitem
\bibitem[Caruso \latin{et~al.}(2016)Caruso, Dauth, {van Setten}, and
  Rinke]{carusoBenchmarkGWApproaches2016}
Caruso,~F.; Dauth,~M.; {van Setten},~M.~J.; Rinke,~P. Benchmark of {{GW
  Approaches}} for the {{GW100 Test Set}}. \emph{J. Chem. Theory Comput.}
  \textbf{2016}, \emph{12}, 5076--5087\relax
\mciteBstWouldAddEndPuncttrue
\mciteSetBstMidEndSepPunct{\mcitedefaultmidpunct}
{\mcitedefaultendpunct}{\mcitedefaultseppunct}\relax
\EndOfBibitem
\bibitem[Grumet \latin{et~al.}(2018)Grumet, Liu, Kaltak, Klime{\v s}, and
  Kresse]{grumetQuasiparticleApproximationFully2018}
Grumet,~M.; Liu,~P.; Kaltak,~M.; Klime{\v s},~J.; Kresse,~G. Beyond the
  Quasiparticle Approximation: {{Fully}} Self-Consistent ${{GW}}$ Calculations.
  \emph{Phys. Rev. B} \textbf{2018}, \emph{98}, 155143\relax
\mciteBstWouldAddEndPuncttrue
\mciteSetBstMidEndSepPunct{\mcitedefaultmidpunct}
{\mcitedefaultendpunct}{\mcitedefaultseppunct}\relax
\EndOfBibitem
\bibitem[Jin \latin{et~al.}(2019)Jin, Su, and
  Yang]{jinRenormalizedSinglesGreen2019}
Jin,~Y.; Su,~N.~Q.; Yang,~W. Renormalized {{Singles Green}}'s {{Function}} for
  {{Quasi-Particle Calculations}} beyond the {{G0W0 Approximation}}. \emph{J.
  Phys. Chem. Lett.} \textbf{2019}, \emph{10}, 447--452\relax
\mciteBstWouldAddEndPuncttrue
\mciteSetBstMidEndSepPunct{\mcitedefaultmidpunct}
{\mcitedefaultendpunct}{\mcitedefaultseppunct}\relax
\EndOfBibitem
\bibitem[Ren \latin{et~al.}(2011)Ren, Tkatchenko, Rinke, and
  Scheffler]{renRandomPhaseApproximationElectron2011}
Ren,~X.; Tkatchenko,~A.; Rinke,~P.; Scheffler,~M. Beyond the {{Random-Phase
  Approximation}} for the {{Electron Correlation Energy}}: {{The Importance}}
  of {{Single Excitations}}. \emph{Phys. Rev. Lett.} \textbf{2011}, \emph{106},
  153003\relax
\mciteBstWouldAddEndPuncttrue
\mciteSetBstMidEndSepPunct{\mcitedefaultmidpunct}
{\mcitedefaultendpunct}{\mcitedefaultseppunct}\relax
\EndOfBibitem
\bibitem[Ren \latin{et~al.}(2013)Ren, Rinke, Scuseria, and
  Scheffler]{renRenormalizedSecondorderPerturbation2013}
Ren,~X.; Rinke,~P.; Scuseria,~G.~E.; Scheffler,~M. Renormalized Second-Order
  Perturbation Theory for the Electron Correlation Energy: {{Concept}},
  Implementation, and Benchmarks. \emph{Phys. Rev. B} \textbf{2013}, \emph{88},
  035120\relax
\mciteBstWouldAddEndPuncttrue
\mciteSetBstMidEndSepPunct{\mcitedefaultmidpunct}
{\mcitedefaultendpunct}{\mcitedefaultseppunct}\relax
\EndOfBibitem
\bibitem[Li \latin{et~al.}(2022)Li, Chen, and
  Yang]{liMultireferenceDensityFunctional2022}
Li,~J.; Chen,~Z.; Yang,~W. Multireference {{Density Functional Theory}} for
  {{Describing Ground}} and {{Excited States}} with {{Renormalized Singles}}.
  \emph{J. Phys. Chem. Lett.} \textbf{2022}, \emph{13}, 894--903\relax
\mciteBstWouldAddEndPuncttrue
\mciteSetBstMidEndSepPunct{\mcitedefaultmidpunct}
{\mcitedefaultendpunct}{\mcitedefaultseppunct}\relax
\EndOfBibitem
\bibitem[Li \latin{et~al.}(2021)Li, Chen, and
  Yang]{liRenormalizedSinglesGreen2021}
Li,~J.; Chen,~Z.; Yang,~W. Renormalized {{Singles Green}}'s {{Function}} in the
  {{T-Matrix Approximation}} for {{Accurate Quasiparticle Energy Calculation}}.
  \emph{J. Phys. Chem. Lett.} \textbf{2021}, \emph{12}, 6203--6210\relax
\mciteBstWouldAddEndPuncttrue
\mciteSetBstMidEndSepPunct{\mcitedefaultmidpunct}
{\mcitedefaultendpunct}{\mcitedefaultseppunct}\relax
\EndOfBibitem
\bibitem[Zhang \latin{et~al.}(2017)Zhang, Su, and
  Yang]{zhangAccurateQuasiparticleSpectra2017}
Zhang,~D.; Su,~N.~Q.; Yang,~W. Accurate {{Quasiparticle Spectra}} from the
  {{T-Matrix Self-Energy}} and the {{Particle}}\textendash{{Particle Random
  Phase Approximation}}. \emph{J. Phys. Chem. Lett.} \textbf{2017}, \emph{8},
  3223--3227\relax
\mciteBstWouldAddEndPuncttrue
\mciteSetBstMidEndSepPunct{\mcitedefaultmidpunct}
{\mcitedefaultendpunct}{\mcitedefaultseppunct}\relax
\EndOfBibitem
\bibitem[Zhang and Yang(2016)Zhang, and
  Yang]{zhangAccurateEfficientCalculation2016}
Zhang,~D.; Yang,~W. Accurate and Efficient Calculation of Excitation Energies
  with the Active-Space Particle-Particle Random Phase Approximation. \emph{J.
  Chem. Phys.} \textbf{2016}, \emph{145}, 144105\relax
\mciteBstWouldAddEndPuncttrue
\mciteSetBstMidEndSepPunct{\mcitedefaultmidpunct}
{\mcitedefaultendpunct}{\mcitedefaultseppunct}\relax
\EndOfBibitem
\bibitem[Pollehn \latin{et~al.}(1998)Pollehn, Schindlmayr, and
  Godby]{pollehnAssessmentLessGreater1998}
Pollehn,~T.~J.; Schindlmayr,~A.; Godby,~R.~W. Assessment of the $GW$
  approximation using Hubbard chains. \emph{J. Phys.: Condens. Matter}
  \textbf{1998}, \emph{10}, 1273--1283\relax
\mciteBstWouldAddEndPuncttrue
\mciteSetBstMidEndSepPunct{\mcitedefaultmidpunct}
{\mcitedefaultendpunct}{\mcitedefaultseppunct}\relax
\EndOfBibitem
\bibitem[Govoni and Galli(2015)Govoni, and Galli]{govoniLargeScaleGW2015}
Govoni,~M.; Galli,~G. Large {{Scale GW Calculations}}. \emph{J. Chem. Theory
  Comput.} \textbf{2015}, \emph{11}, 2680--2696\relax
\mciteBstWouldAddEndPuncttrue
\mciteSetBstMidEndSepPunct{\mcitedefaultmidpunct}
{\mcitedefaultendpunct}{\mcitedefaultseppunct}\relax
\EndOfBibitem
\bibitem[Lee \latin{et~al.}(1999)Lee, Gunnarsson, and
  Hedin]{leeTransitionAdiabaticSudden1999}
Lee,~J.~D.; Gunnarsson,~O.; Hedin,~L. Transition from the Adiabatic to the
  Sudden Limit in Core-Level Photoemission: {{A}} Model Study of a Localized
  System. \emph{Phys. Rev. B} \textbf{1999}, \emph{60}, 8034--8049\relax
\mciteBstWouldAddEndPuncttrue
\mciteSetBstMidEndSepPunct{\mcitedefaultmidpunct}
{\mcitedefaultendpunct}{\mcitedefaultseppunct}\relax
\EndOfBibitem
\bibitem[Rinke \latin{et~al.}(2005)Rinke, Qteish, Neugebauer, Freysoldt, and
  Scheffler]{rinkeCombiningGWcalculationsExactexchangeDensityfunctional2005}
Rinke,~P.; Qteish,~A.; Neugebauer,~J.; Freysoldt,~C.; Scheffler,~M.
  {{CombiningGWcalculations}} with Exact-Exchange Density-Functional Theory: An
  Analysis of Valence-Band Photoemission for Compound Semiconductors. \emph{New
  J. Phys.} \textbf{2005}, \emph{7}, 126--126\relax
\mciteBstWouldAddEndPuncttrue
\mciteSetBstMidEndSepPunct{\mcitedefaultmidpunct}
{\mcitedefaultendpunct}{\mcitedefaultseppunct}\relax
\EndOfBibitem
\bibitem[Ren \latin{et~al.}(2012)Ren, Rinke, Joas, and
  Scheffler]{renRandomphaseApproximationIts2012}
Ren,~X.; Rinke,~P.; Joas,~C.; Scheffler,~M. Random-Phase Approximation and Its
  Applications in Computational Chemistry and Materials Science. \emph{J Mater
  Sci} \textbf{2012}, \emph{47}, 7447--7471\relax
\mciteBstWouldAddEndPuncttrue
\mciteSetBstMidEndSepPunct{\mcitedefaultmidpunct}
{\mcitedefaultendpunct}{\mcitedefaultseppunct}\relax
\EndOfBibitem
\bibitem[Szabo and Ostlund(2012)Szabo, and
  Ostlund]{szaboModernQuantumChemistry2012}
Szabo,~A.; Ostlund,~N.~S. \emph{Modern {{Quantum Chemistry}}: {{Introduction}}
  to {{Advanced Electronic Structure Theory}}}; {Courier Corporation},
  2012\relax
\mciteBstWouldAddEndPuncttrue
\mciteSetBstMidEndSepPunct{\mcitedefaultmidpunct}
{\mcitedefaultendpunct}{\mcitedefaultseppunct}\relax
\EndOfBibitem
\bibitem[Blum \latin{et~al.}(2009)Blum, Gehrke, Hanke, Havu, Havu, Ren, Reuter,
  and Scheffler]{blumInitioMolecularSimulations2009}
Blum,~V.; Gehrke,~R.; Hanke,~F.; Havu,~P.; Havu,~V.; Ren,~X.; Reuter,~K.;
  Scheffler,~M. Ab Initio Molecular Simulations with Numeric Atom-Centered
  Orbitals. \emph{Comput. Phys. Commun} \textbf{2009}, \emph{180},
  2175--2196\relax
\mciteBstWouldAddEndPuncttrue
\mciteSetBstMidEndSepPunct{\mcitedefaultmidpunct}
{\mcitedefaultendpunct}{\mcitedefaultseppunct}\relax
\EndOfBibitem
\bibitem[Havu \latin{et~al.}(2009)Havu, Blum, Havu, and
  Scheffler]{havuEfficientIntegrationAllelectron2009}
Havu,~V.; Blum,~V.; Havu,~P.; Scheffler,~M. Efficient {{O}}({{N}}) Integration
  for All-Electron Electronic Structure Calculation Using Numeric Basis
  Functions. \emph{J. Comput. Phys.} \textbf{2009}, \emph{228},
  8367--8379\relax
\mciteBstWouldAddEndPuncttrue
\mciteSetBstMidEndSepPunct{\mcitedefaultmidpunct}
{\mcitedefaultendpunct}{\mcitedefaultseppunct}\relax
\EndOfBibitem
\bibitem[qm4()]{qm4d}
See http://www.qm4d.info for an in-house program for QM/MM simulations\relax
\mciteBstWouldAddEndPuncttrue
\mciteSetBstMidEndSepPunct{\mcitedefaultmidpunct}
{\mcitedefaultendpunct}{\mcitedefaultseppunct}\relax
\EndOfBibitem
\bibitem[Perdew \latin{et~al.}(1996)Perdew, Burke, and
  Ernzerhof]{perdewGeneralizedGradientApproximation1996}
Perdew,~J.~P.; Burke,~K.; Ernzerhof,~M. Generalized {{Gradient Approximation
  Made Simple}}. \emph{Phys. Rev. Lett.} \textbf{1996}, \emph{77},
  3865--3868\relax
\mciteBstWouldAddEndPuncttrue
\mciteSetBstMidEndSepPunct{\mcitedefaultmidpunct}
{\mcitedefaultendpunct}{\mcitedefaultseppunct}\relax
\EndOfBibitem
\bibitem[Atalla \latin{et~al.}(2013)Atalla, Yoon, Caruso, Rinke, and
  Scheffler]{atallaHybridDensityFunctional2013}
Atalla,~V.; Yoon,~M.; Caruso,~F.; Rinke,~P.; Scheffler,~M. Hybrid Density
  Functional Theory Meets Quasiparticle Calculations: {{A}} Consistent
  Electronic Structure Approach. \emph{Phys. Rev. B} \textbf{2013}, \emph{88},
  165122\relax
\mciteBstWouldAddEndPuncttrue
\mciteSetBstMidEndSepPunct{\mcitedefaultmidpunct}
{\mcitedefaultendpunct}{\mcitedefaultseppunct}\relax
\EndOfBibitem
\bibitem[Adamo and Barone(1999)Adamo, and
  Barone]{adamoReliableDensityFunctional1999}
Adamo,~C.; Barone,~V. Toward Reliable Density Functional Methods without
  Adjustable Parameters: {{The PBE0}} Model. \emph{J. Chem. Phys.}
  \textbf{1999}, \emph{110}, 6158--6170\relax
\mciteBstWouldAddEndPuncttrue
\mciteSetBstMidEndSepPunct{\mcitedefaultmidpunct}
{\mcitedefaultendpunct}{\mcitedefaultseppunct}\relax
\EndOfBibitem
\bibitem[Ernzerhof and Scuseria(1999)Ernzerhof, and
  Scuseria]{ernzerhofAssessmentPerdewBurke1999}
Ernzerhof,~M.; Scuseria,~G.~E. Assessment of the
  {{Perdew}}\textendash{{Burke}}\textendash{{Ernzerhof}} Exchange-Correlation
  Functional. \emph{J. Chem. Phys.} \textbf{1999}, \emph{110}, 5029--5036\relax
\mciteBstWouldAddEndPuncttrue
\mciteSetBstMidEndSepPunct{\mcitedefaultmidpunct}
{\mcitedefaultendpunct}{\mcitedefaultseppunct}\relax
\EndOfBibitem
\bibitem[Becke(1993)]{beckeDensityFunctionalThermochemistry1993}
Becke,~A.~D. Density-functional Thermochemistry. {{III}}. {{The}} Role of Exact
  Exchange. \emph{J. Chem. Phys.} \textbf{1993}, \emph{98}, 5648--5652\relax
\mciteBstWouldAddEndPuncttrue
\mciteSetBstMidEndSepPunct{\mcitedefaultmidpunct}
{\mcitedefaultendpunct}{\mcitedefaultseppunct}\relax
\EndOfBibitem
\bibitem[Lee \latin{et~al.}(1988)Lee, Yang, and
  Parr]{leeDevelopmentColleSalvettiCorrelationenergy1988}
Lee,~C.; Yang,~W.; Parr,~R.~G. Development of the {{Colle-Salvetti}}
  Correlation-Energy Formula into a Functional of the Electron Density.
  \emph{Phys. Rev. B} \textbf{1988}, \emph{37}, 785--789\relax
\mciteBstWouldAddEndPuncttrue
\mciteSetBstMidEndSepPunct{\mcitedefaultmidpunct}
{\mcitedefaultendpunct}{\mcitedefaultseppunct}\relax
\EndOfBibitem
\bibitem[Dunning(1989)]{dunningGaussianBasisSets1989}
Dunning,~T.~H. Gaussian Basis Sets for Use in Correlated Molecular
  Calculations. {{I}}. {{The}} Atoms Boron through Neon and Hydrogen. \emph{J.
  Chem. Phys.} \textbf{1989}, \emph{90}, 1007--1023\relax
\mciteBstWouldAddEndPuncttrue
\mciteSetBstMidEndSepPunct{\mcitedefaultmidpunct}
{\mcitedefaultendpunct}{\mcitedefaultseppunct}\relax
\EndOfBibitem
\bibitem[Wilson \latin{et~al.}(1996)Wilson, {van Mourik}, and
  Dunning]{wilsonGaussianBasisSets1996}
Wilson,~A.~K.; {van Mourik},~T.; Dunning,~T.~H. Gaussian Basis Sets for Use in
  Correlated Molecular Calculations. {{VI}}. {{Sextuple}} Zeta Correlation
  Consistent Basis Sets for Boron through Neon. \emph{J. Mol. Struct. THEOCHEM}
  \textbf{1996}, \emph{388}, 339--349\relax
\mciteBstWouldAddEndPuncttrue
\mciteSetBstMidEndSepPunct{\mcitedefaultmidpunct}
{\mcitedefaultendpunct}{\mcitedefaultseppunct}\relax
\EndOfBibitem
\bibitem[Vahtras \latin{et~al.}(1993)Vahtras, Alml{\"o}f, and
  Feyereisen]{vahtrasIntegralApproximationsLCAOSCF1993}
Vahtras,~O.; Alml{\"o}f,~J.; Feyereisen,~M.~W. Integral Approximations for
  {{LCAO-SCF}} Calculations. \emph{Chem. Phys. Lett} \textbf{1993}, \emph{213},
  514--518\relax
\mciteBstWouldAddEndPuncttrue
\mciteSetBstMidEndSepPunct{\mcitedefaultmidpunct}
{\mcitedefaultendpunct}{\mcitedefaultseppunct}\relax
\EndOfBibitem
\bibitem[Weigend \latin{et~al.}(2002)Weigend, K{\"o}hn, and
  H{\"a}ttig]{weigendEfficientUseCorrelation2002}
Weigend,~F.; K{\"o}hn,~A.; H{\"a}ttig,~C. Efficient Use of the Correlation
  Consistent Basis Sets in Resolution of the Identity {{MP2}} Calculations.
  \emph{J. Chem. Phys.} \textbf{2002}, \emph{116}, 3175--3183\relax
\mciteBstWouldAddEndPuncttrue
\mciteSetBstMidEndSepPunct{\mcitedefaultmidpunct}
{\mcitedefaultendpunct}{\mcitedefaultseppunct}\relax
\EndOfBibitem
\bibitem[Sankari \latin{et~al.}(2006)Sankari, Ehara, Nakatsuji, De~Fanis,
  Aksela, Sorensen, Piancastelli, Kukk, and Ueda]{sankariHighResolution1s2006}
Sankari,~R.; Ehara,~M.; Nakatsuji,~H.; De~Fanis,~A.; Aksela,~H.;
  Sorensen,~S.~L.; Piancastelli,~M.~N.; Kukk,~E.; Ueda,~K. High Resolution
  {{O}} 1s Photoelectron Shake-up Satellite Spectrum of {{H2O}}. \emph{Chem.
  Phys. Lett} \textbf{2006}, \emph{422}, 51--57\relax
\mciteBstWouldAddEndPuncttrue
\mciteSetBstMidEndSepPunct{\mcitedefaultmidpunct}
{\mcitedefaultendpunct}{\mcitedefaultseppunct}\relax
\EndOfBibitem
\bibitem[Schirmer \latin{et~al.}(1987)Schirmer, Angonoa, Svensson, Nordfors,
  and Gelius]{schirmerHighenergyPhotoelectron1s1987}
Schirmer,~J.; Angonoa,~G.; Svensson,~S.; Nordfors,~D.; Gelius,~U. High-Energy
  Photoelectron {{C}} 1s and {{O}} 1s Shake-up Spectra of {{CO}}. \emph{J.
  Phys. B: Atom. Mol. Phys.} \textbf{1987}, \emph{20}, 6031--6040\relax
\mciteBstWouldAddEndPuncttrue
\mciteSetBstMidEndSepPunct{\mcitedefaultmidpunct}
{\mcitedefaultendpunct}{\mcitedefaultseppunct}\relax
\EndOfBibitem
\bibitem[Shishkin and Kresse(2007)Shishkin, and
  Kresse]{shishkinSelfconsistentGWCalculations2007}
Shishkin,~M.; Kresse,~G. Self-Consistent ${{GW}}$ Calculations for
  Semiconductors and Insulators. \emph{Phys. Rev. B} \textbf{2007}, \emph{75},
  235102\relax
\mciteBstWouldAddEndPuncttrue
\mciteSetBstMidEndSepPunct{\mcitedefaultmidpunct}
{\mcitedefaultendpunct}{\mcitedefaultseppunct}\relax
\EndOfBibitem
\bibitem[Byun and {\"O}{\u g}{\"u}t(2019)Byun, and {\"O}{\u
  g}{\"u}t]{byunPracticalGWScheme2019}
Byun,~Y.-M.; {\"O}{\u g}{\"u}t,~S. Practical {{GW}} Scheme for Electronic
  Structure of 3d-Transition-Metal Monoxide Anions: {{ScO}}-, {{TiO}}-,
  {{CuO}}-, and {{ZnO}}-. \emph{J. Chem. Phys.} \textbf{2019}, \emph{151},
  134305\relax
\mciteBstWouldAddEndPuncttrue
\mciteSetBstMidEndSepPunct{\mcitedefaultmidpunct}
{\mcitedefaultendpunct}{\mcitedefaultseppunct}\relax
\EndOfBibitem
\bibitem[Marom \latin{et~al.}(2012)Marom, Caruso, Ren, Hofmann,
  K{\"o}rzd{\"o}rfer, Chelikowsky, Rubio, Scheffler, and
  Rinke]{maromBenchmarkGWMethods2012}
Marom,~N.; Caruso,~F.; Ren,~X.; Hofmann,~O.~T.; K{\"o}rzd{\"o}rfer,~T.;
  Chelikowsky,~J.~R.; Rubio,~A.; Scheffler,~M.; Rinke,~P. Benchmark of {{GW}}
  Methods for Azabenzenes. \emph{Phys. Rev. B} \textbf{2012}, \emph{86},
  245127\relax
\mciteBstWouldAddEndPuncttrue
\mciteSetBstMidEndSepPunct{\mcitedefaultmidpunct}
{\mcitedefaultendpunct}{\mcitedefaultseppunct}\relax
\EndOfBibitem
\bibitem[Pireaux \latin{et~al.}(1976)Pireaux, Svensson, Basilier, Malmqvist,
  Gelius, Caudano, and Siegbahn]{pireauxCoreelectronRelaxationEnergies1976}
Pireaux,~J.~J.; Svensson,~S.; Basilier,~E.; Malmqvist,~P.-{\AA}.; Gelius,~U.;
  Caudano,~R.; Siegbahn,~K. Core-Electron Relaxation Energies and Valence-Band
  Formation of Linear Alkanes Studied in the Gas Phase by Means of Electron
  Spectroscopy. \emph{Phys. Rev. A} \textbf{1976}, \emph{14}, 2133--2145\relax
\mciteBstWouldAddEndPuncttrue
\mciteSetBstMidEndSepPunct{\mcitedefaultmidpunct}
{\mcitedefaultendpunct}{\mcitedefaultseppunct}\relax
\EndOfBibitem
\bibitem[Travnikova \latin{et~al.}(2012)Travnikova, B{\o}rve, Patanen,
  S{\"o}derstr{\"o}m, Miron, S{\ae}thre, M{\aa}rtensson, and
  Svensson]{travnikovaESCAMoleculeHistorical2012}
Travnikova,~O.; B{\o}rve,~K.~J.; Patanen,~M.; S{\"o}derstr{\"o}m,~J.;
  Miron,~C.; S{\ae}thre,~L.~J.; M{\aa}rtensson,~N.; Svensson,~S. The {{ESCA}}
  Molecule\textemdash{{Historical}} Remarks and New Results. \emph{J. Electron
  Spectrosc. Relat. Phenom.} \textbf{2012}, \emph{185}, 191--197\relax
\mciteBstWouldAddEndPuncttrue
\mciteSetBstMidEndSepPunct{\mcitedefaultmidpunct}
{\mcitedefaultendpunct}{\mcitedefaultseppunct}\relax
\EndOfBibitem
\bibitem[Gelius \latin{et~al.}(1973)Gelius, Basilier, Svensson, Bergmark, and
  Siegbahn]{geliusHighResolutionESCA1973}
Gelius,~U.; Basilier,~E.; Svensson,~S.; Bergmark,~T.; Siegbahn,~K. A High
  Resolution {{ESCA}} Instrument with {{X-ray}} Monochromator for Gases and
  Solids. \emph{J. Electron Spectrosc. Relat. Phenom.} \textbf{1973}, \emph{2},
  405--434\relax
\mciteBstWouldAddEndPuncttrue
\mciteSetBstMidEndSepPunct{\mcitedefaultmidpunct}
{\mcitedefaultendpunct}{\mcitedefaultseppunct}\relax
\EndOfBibitem
\bibitem[Delesma \latin{et~al.}(2018)Delesma, {Van den Bossche}, Gr{\"o}nbeck,
  Calaminici, K{\"o}ster, and Pettersson]{delesmaChemicalViewXray2018}
Delesma,~F.~A.; {Van den Bossche},~M.; Gr{\"o}nbeck,~H.; Calaminici,~P.;
  K{\"o}ster,~A.~M.; Pettersson,~L. G.~M. A {{Chemical View}} on {{X-ray
  Photoelectron Spectroscopy}}: The {{ESCA Molecule}} and {{Surface-to-Bulk XPS
  Shifts}}. \emph{ChemPhysChem} \textbf{2018}, \emph{19}, 169--174\relax
\mciteBstWouldAddEndPuncttrue
\mciteSetBstMidEndSepPunct{\mcitedefaultmidpunct}
{\mcitedefaultendpunct}{\mcitedefaultseppunct}\relax
\EndOfBibitem
\bibitem[{Van den Bossche} \latin{et~al.}(2014){Van den Bossche}, Martin,
  Gustafson, Hakanoglu, Weaver, Lundgren, and
  Gr{\"o}nbeck]{vandenbosscheEffectsNonlocalExchange2014}
{Van den Bossche},~M.; Martin,~N.~M.; Gustafson,~J.; Hakanoglu,~C.;
  Weaver,~J.~F.; Lundgren,~E.; Gr{\"o}nbeck,~H. Effects of Non-Local Exchange
  on Core Level Shifts for Gas-Phase and Adsorbed Molecules. \emph{J. Chem.
  Phys.} \textbf{2014}, \emph{141}, 034706\relax
\mciteBstWouldAddEndPuncttrue
\mciteSetBstMidEndSepPunct{\mcitedefaultmidpunct}
{\mcitedefaultendpunct}{\mcitedefaultseppunct}\relax
\EndOfBibitem
\bibitem[Travnikova \latin{et~al.}(2019)Travnikova, Patanen,
  S{\"o}derstr{\"o}m, Lindblad, Kas, Vila, C{\'e}olin, Marchenko, Goldsztejn,
  Guillemin, Journel, Carroll, B{\o}rve, Decleva, Rehr, M{\aa}rtensson, Simon,
  Svensson, and S{\ae}thre]{travnikovaEnergyDependentRelativeCross2019}
Travnikova,~O. \latin{et~al.}  Energy-{{Dependent Relative Cross Sections}} in
  {{Carbon}} 1s {{Photoionization}}: {{Separation}} of {{Direct Shake}} and
  {{Inelastic Scattering Effects}} in {{Single Molecules}}. \emph{J. Phys.
  Chem. A} \textbf{2019}, \emph{123}, 7619--7636\relax
\mciteBstWouldAddEndPuncttrue
\mciteSetBstMidEndSepPunct{\mcitedefaultmidpunct}
{\mcitedefaultendpunct}{\mcitedefaultseppunct}\relax
\EndOfBibitem
\bibitem[Golze(2022)]{nomad_repo}
Golze,~D. Dataset in {NOMAD} repository. 2022; DOI will be added after
  revision\relax
\mciteBstWouldAddEndPuncttrue
\mciteSetBstMidEndSepPunct{\mcitedefaultmidpunct}
{\mcitedefaultendpunct}{\mcitedefaultseppunct}\relax
\EndOfBibitem
\end{mcitethebibliography}

\end{document}